\pdfoutput=1
\documentclass[12pt,preprint]{aastex}
\usepackage{color}

\shorttitle{Shock-Cloud interaction}
\shortauthors{Takahashi.}

\begin{document}
\title{Shock-Cloud Interaction in the Solar Corona}

\author{
Takuya Takahashi\altaffilmark{1,2}
}

\email{takahashi@kwasan.kyoto-u.ac.jp}
\altaffiltext{1}{
Department of Astronomy, Kyoto University,
Sakyo, Kyoto, 606-8502, Japan}
\altaffiltext{2}{
Kwasan and Hida Observatories, Kyoto University,
Yamashina, Kyoto 607-8471, Japan.}

\begin{abstract}
Flare associated coronal shock waves sometimes interact with solar prominences leading to large amplitude prominence oscillations. Such prominence activation gives us unique opportunity to track time evolution of shock-cloud interaction in cosmic plasmas. Although the dynamics of interstellar shock-cloud interaction is extensively studied, coronal shock-solar prominence interaction is rarely studied in the context of shock-cloud interaction. 
Associated with X5.4 class solar flare occurred on 7 March, 2012, a globally propagated coronal shock wave interacted with a polar prominence leading to large amplitude prominence oscillation. In this paper, we studied bulk acceleration and excitation of internal flow of the shocked prominence using three-dimensional MHD simulations. We studied eight magnetohydrodynamic (MHD) simulation runs with different mass density structure of the prominence, and one hydrodynamic simulation run, and compared the result. In order to compare observed motion of activated  prominence with corresponding simulation, we also studied prominence activation by injection of triangular shaped coronal shock. We found that magnetic tension force mainly accelerate (and then decelerate) the prominence. The internal flow, on the other hand, is excited during the shock front sweeps through the the prominence and damps almost exponentially. We construct phenomenological model of bulk momentum transfer from shock to the prominence, which agreed quantitatively with all the simulation results. Based on the phenomenological prominence-activation model, we diagnosed physical parameters of coronal shock wave. The estimated energy of the coronal shock is several percent of total energy released during the X5.4 flare.
\end{abstract}

\keywords{magnetohydrodynamics(MHD) --- Shock waves --- 
Sun: corona --- Sun: filaments, prominences --- Sun: flares --- ISM: clouds}

\section{Introduction}
The interaction between interstellar clouds and shock waves associated with for example supernova remnants, H II regions or stellar winds has been studied as one of the most fundamental problem of interstellar gas dynamics. Shock-molecular cloud (MC) interaction is especially important as a process that dynamically drives star formation. Studies using hydrodynamics or magnetohydrodynamics simulations revealed that strong interstellar shock wave injected into the molecular cloud crush and destroy it through shock injection and ensuing hydrodynamical instabilities such as Kelvin-Helmholtz (KH), Rayleigh-Taylor (RT) and Richtmyer-Meshkov (RM) instabilities, although inclusion of magnetic field strongly suppress those instabilities \citep{wood1976, nitt1982, klei1994, macl1994}. Magnetic field orientation and mass density structure within clouds also affects significantly the later phase of the dynamics of molecular cloud impacted by shock wave \citep{polu2002, patn2005, shin2008}. Recent observations and numerical simulations reveal multi-phase multi-scale dynamics of filamentary clouds - shock interaction within a cloud is important for the evolution of star forming molecular clouds with the help of thermal or self gravitational instabilities \citep{inou2012, doba2014, mats2015}.

On the other hand, shock waves are frequently observed in the corona of the Sun \citep{more1960,uchi1968,thom2000,warm2001,vrsn2002,liu2010}. In the corona, magnetic field reconnection allows magnetic field energy released in a catastrophic way resulting in largest explosion in the solar system. This is called solar flares. During solar flares, typically $10^{29}$ - $10^{33} $ergs of magnetic field energy stored in the corona is converted to thermal, kinetic, radiation and high energy particle kinetic energy in a short (several minutes) time scale \citep{shibm2011}.

As a result of sudden energy release in solar flares, a part of energy propagate globally in the corona as a form of non linear fast mode MHD wave (or shock).  Recently, high cadence extreme ultraviolet (EUV) observation of the solar corona by atmospheric imaging assembly (AIA; \citet{tit2006, lem2011}) on board solar dynamics observatory ({\it SDO}; Pesnell et al. 2012) allows detailed imaging observation of such global shock waves in the corona \citep{asai2012}. These flare associated shock waves in the lower solar corona are thought to be weak fast mode MHD shocks \citep{ma2011,gopa2012}. Plasma ejections (coronal mass ejections, CMEs) associated with solar flares propagate in the interplanetary space, sometimes with clear shock fronts at their noses \citep{wang2001}. These propagating shockwaves in the upper corona or interplanetary space are observed in coronagraph images or in radio dynamic spectrum observations \citep{kai1970}. They are also observed as a sudden change in plasma parameters of solar wind velocity, temperature, density and magnetic field strength observed in-situ in front of the Earth \citep{wang2001}.

We also have 'clouds' in the solar corona, the 'prominences'. They are cool and dense plasma floating within hot and rarefied corona. They are supported by magnetic tension force against gravity\citep{lab2010, mac2010}. Recent high resolution and high sensitivity observation by Solar Optical Telescope (SOT; Tsuneta et al. 2008) on board $Hinode$ satellite\citep{kosu2007} revealed highly dynamic nature of solar prominence, with continuous oscillation and turbulent flow\citep{berg2008,berg2011}. They give us rare opportunity for studying dynamics of partially ionized plasma in detail, whose plasma parameters are very difficult to reach in laboratory or in other space objects. The excitation mechanism of such chaotic flows within prominence material are also discussed recently. \citet{hill2013} discussed photospheric motion as one possible mechanism that drive long frequency small amplitude prominence oscillations. Non-linear MHD waves propagating upward around prominence foot is studied by \citep{offm2015}. The Magnetic Rayleigh-Taylor instability invoked by interchange reconnection between magnetic field lines supporting prominence plasmas is discussed as an excitation mechanism for multiple plume-like upflows observed with {\it Hinode}/SOT which help mix up prominence plasmas \citep{hill2012}. Magnetic Kelvin-Helmholtz instability excited near the absorption layer of transversely oscillating prominence in the corona helps cascade energy and heat the prominence plasma through turbulence excitation and current sheet formation \citep{okam2015, anto2015}.

Sometimes, coronal shock waves associated with flares hit solar prominences and lead to excitement of large amplitude prominence oscillations. These oscillations give us information of physical properties of prominences such as magnetic field strength, density, and eruptive stability and have been studied widely \citep{isob2007,gilb2008}.

In contrast to interstellar shock-molecular cloud interaction that has been studied widely, the excitation process of large amplitude solar prominences or prominence activations by coronal shock injection has not been studied in detail in the context of shock-cloud interaction. In contrast to the situation in interstellar medium where strong shock wave often interact with molecular clouds, the prominence activation is the interaction between prominence and weak fast mode MHD shock wave with fast mode Mach number being between 1.1 and 1.5 \citep{naru2002, gre2011, taka2015}. The time scale of shock-prominence interaction  is several minutes, which offers us unique opportunity to study in detail the time evolution of shock-cloud interaction process in cosmic plasmas.

In this paper, we analyze observational data of shock-prominence interaction obtained by SDO/AIA and compare them with numerical MHD simulation. Especially, we focus quantitatively on how wave momentum is transferred to cloud material through shock injection, flow drag and magnetic tension acceleration processes with 3D MHD simulation. We also discuss the effect of internal structure such as volume filling factor on chaotic flow excitation and its damping. We compare the MHD simulation results with hydrodynamic simulation and discuss the role of magnetic field in bulk prominence acceleration as well as excitement and damping of internal flows. We make a phenomenological model that describe momentum transfer from coronal shock to the prominence. We validate the phenomenological model by comparison with simulation, and apply the phenomenological model of prominence activation in the context of diagnosing coronal shock properties. Lastly, we compare shock-prominence interaction with interstellar shock-molecular cloud interaction in the context of MHD shock-cloud interaction.

\section{EUV observation of a prominence activation by coronal shock wave}
On March 7, 2012, an X5.4 class flare occurred at NOAA active region (AR) 11429 located at north east quadrant of the solar disk. The soft X-ray light curve obtained by GOES peaked at 00:24UT on March 7(Figure 1(a)). This flare is the second largest one in current solar cycle(cycle 24). The flare was associated with very fast CME whose velocity was about 2684km/s.\footnote{see,
http://cdaw.gsfc.nasa.gov/CME\_list/UNIVERSAL/2012\_03/univ2012\_03.html} Figure 1 (b) is the composite of coronagraph images obtained by {\it SOHO}/LASCO C2 and EUV image obtained by {\it SDO}/AIA 193 $~{\AA}$ band both taken at 00:36UT. We can see the shock front surrounding the CME ejecta in {\it SOHO}/LASCO C2 image(Figure 1(b)). Figure 1(d) shows {\it SDO}/AIA 193$~{\AA}$ difference image at 00:18UT. We see a dome-like bright structure expanding above AR 11429 in Figure 1(d).

\citet{taka2015} estimated the propagation speed of the leading shock front ahead of the CME as 672km s$^{-1}$ based on the analysis of dynamic spectrum obtained with the Hiraiso Radio Spectrograph (HiRAS, Kondo et al. 1995). It seems strange that the estimated speed of the leading shock front ahead of the CME is much slower than the CME speed estimated with {\it SOHO}/LASCO. We looked at the coronagraph observation data by {\it STEREO-B}/COR1 and radio dynamic spectrum once again to check the consistency. The hight of the leading edge of the CME ejecta seen in {\it STEREO-B}/COR1 image taken at 00:26UT is larger than 2 $R_s$ measured from the photosphere, where $R_s$ is the solar radius. On the other hand, the radio dynamic spectrum during the flare period is complicated and composed of Type II, Type IV and possibly Type III bursts. We noticed \citet{taka2015} mistook type IV burst for type II burst, resulting in the CME speed estimation inconsistent with coronagraph observation. From 00:17:10UT (indicated as $P_1$ in figure 2) to 00:17:38UT (indicated as $P_2$ in Figure 2), we see a clear linear structure in the radio dynamic spectrum showing the characteristic signature of Type II burst, whose frequency drifted from the 112MHz to 88MHz during the period. Assuming the type II burst signature corresponds to the first harmonics of the plasma oscillation at the upstream of the leading shock front, we got the propagation speed of 1.9$\times$10$^3$ km s$^{-1}$ based on \citep{new1961,mann1999}, which seems to be consistent with coronagraph observations.

The footprints of the dome-like shock front which propagated to the north in the lower corona was especially bright in AIA 193 $~{\AA}$ images. That shock front propagated further to the north and hit a prominence located at the north pole. The interaction between the shock wave and the polar prominence resulted in large amplitude prominence oscillation (LAPO). We call the excitation process of LAPO, 'prominence activation' further on. Figure 3 shows the time evolution of the prominence activation seen in AIA 193$~{\AA}$ and 304$~{\AA}$ images. The FOV of Figure 3 is shown as the black rectangle in Figure1 (c). 

The polar prominence was seen bright in AIA 304 $~{\AA}$ images while it was seen dark in AIA 193 images. Figure 3 (a) and (b) show the prominence just before it is activated by the shock. Figure 3 (c) and (d) show the prominence just hit by the coronal shock front. The shocked part of the prominence seen in AIA 304 {\AA} image became about twice as bright as its original brightness. The bright part propagated in the direction of shock propagation and at the same time the prominence started to move in the shock propagation direction. Figure 3 (i) and (j) show time-distance diagrams along the cut AB shown in Figure 3 (a) between 00:20UT and 01:30UT. In figure 3 (i), the coronal shock front is seen as bright propagating structure. The propagation speed of the coronal shock front in the plane of the sky is measured to be 380~km~s~$^{-1}$ from Figure 3 (i). When the coronal shock front reached the dark prominence, the prominence was abruptly accelerated. The initial activated prominence speed was 48~km~s~$^{-1}$ measured from Figure 3 (i). We can clearly see the sudden brightening of the prominence during its activation in Figure 3 (j). In Figure 3 (j), we can see somewhat chaotic movement of prominence threads during LAPO. We note here that we have neglected the line-of-sight component of the speed of shock propagation and activated prominence, so the estimated shock speed and prominence velocity should be regarded as lower limits. Figure 3 (g) and (h) show the prominence when its displacement from the original position was largest. Note that the time scale of prominence activation is several minutes in this event while the period of LAPO is longer than an hour. The white rectangle in Figure 3 (i) corresponds to the prominence activation process that is also shown in figure 20 (a).

\section{3D MHD simulation of Prominence activation}
 In order to study in detail the physics of prominence activation, we carried out three-dimensional MHD simulation.

\subsection{Numerical methods}
We numerically solved the following resistive MHD equations,
\begin{equation}
\frac{\partial\rho}{\partial t}=-\nabla\cdot\left(\rho{\bf V}\right)
\end{equation}
\begin{equation}
\rho\frac{\partial {\bf V}}{\partial t}=-\rho{\bf V}\cdot\nabla{\bf V}-\nabla p+{\bf J}\times{\bf B}
\end{equation}
\begin{equation}
\frac{\partial p}{\partial t}=-\nabla\cdot\left({\bf V}p\right)-\left(\gamma-1\right)\left(p\nabla\cdot{\bf V}-\eta_0{\bf J}^2\right)
\end{equation}
\begin{equation}
\frac{\partial {\bf B} }{\partial t}-\nabla\times\biggl({\bf V}\times{\bf B}-\eta_0{\bf J}\biggr)=-\nabla\psi
\end{equation}
\begin{equation}
\frac{\partial \psi}{\partial t}+c_h^2\nabla\cdot{\bf B}+\frac{c_h^2}{c_p^2}\psi=0
\end{equation}
, where $\rho,~p,~\bf{B}$ and $\bf{V}$ are mass density, gas pressure, magnetic field and velocity, respectively. ${\bf J}=\nabla \times {\bf B}$ is the electrical current density. $\eta_0$ is uniform electrical resistivity and $\gamma=5/3$ is specific heat ratio.
In the induction equation, additional variable $\psi$ is introduced in order to remove numerical $\nabla\cdot{\bf B}$ as proposed by\citep{dedn2002}.
The numerical scheme we used is Harten-Lax-van Leer-Discontinuities (HLLD) approximate Riemann solver \citep{miyo2005} with second-order total variation diminishing (TVD) Monotonic Upstream-Centered Scheme for Conservation Laws (MUSCL) and second order Runge-Kutta time integration.

\subsection{Initial conditions}
We studied eight simulation cases of coronal shock-prominence interaction, with different density structure of the prominence for each run. In Run $A_1, A_2, A_3$ and $A_4$, the prominence is a uniformly dense spherical plasma of radius $R_p$. Initially, the center of the spherical prominence was at $(x, y, z) = ( -R_p, 0, 0)$. In this model, $R_p=3$ is used. The whole prominence region is where $r_p=\sqrt{(x+R_p)^2+y^2+z^2}<R_p$. 
We modeled the prominence as a sphere of plasma in stead of a cylindrical plasma in this paper in order to also study the effect of magnetic tension force induced by shock injection which is revealed to be important in prominence activation as discussed in section 3.4. When the time scale of prominence activation is much smaller than that of LAPO (as in the case of the event discussed in section 2), the global magnetic field structure will not play a significant role on the dynamics of prominence activation process. In that case we can separate the physics that govern prominence activation with that govern the ensued LAPO. In order to focus only on the physics of prominence activation, we think of such a situation in this study. We set initial magnetic field to be uniform, neglecting the effect of global loop curvature and gravity. As a result, gas pressure also become uniform due to total pressure balance between corona and prominence.
The coronal shock front was initially at $x= -2R_p$. The fast mode shock wave propagate in the positive $x$ direction with density compression ratio $r_{sh,cor}=1.37$. The initial magnetic field is in z-direction, so the injected fast mode shock is perpendicular one. We note that in reality, oblique components may play roles in prominence activation, especially in the plane perpendicular to the shock propagation direction. We do not take into account such oblique dynamics induced by oblique shocks and focus only on the dynamics of activated prominence in the direction of shock propagation, and discuss the effect of prominence mass to the activation dynamics.

In Run $B_1, B_2, B_3$ and $B_4$, the prominence consists of 200 randomly sized spherical clumps of spherical shape put inside the whole sphere of radius $R_p$ (Figure 4(a) and (b)). From the comparison of Runs $B_i$ with Runs $A_i$, we study the role of internal density structure of the prominence in its bulk acceleration and excitation of internal chaotic flow, and also see how well uniform prominence approximation works. In reality, we only solved the numerical box with positive y and z coordinate in order to reduce the numerical cost (i.e. motion of only $\sim 50$ clumps are simulated). From here, we express the density distribution of prominence as $\rho_p$ for convenience.

In Runs $B_i$, clumps are distributed so that the average mass density within the prominence region ($r_p<R_p$) become $\overline{\rho_{p,i}}=f_i\rho_{clump}+(1-f_i)\rho_{cor}$, where $\rho_{cor}$ is the mass density of background (shock upstream) corona and $\rho_{clump}=100\rho_{cor}$ is the mass density of each clump. The volume filling factors $f_i$ are set to be ($f_1, f_2, f_3, f_4)=(0.05, 0.1, 0.3, 1)$. In Runs $A_i$, the mass density of the prominence is uniformly set to be $\rho_p=f_i\rho_{clump}+(1-f_i)\rho_{cor}$ (i.e. both $A_i$ and $B_i$ prominences have the same average density over the entire volume. Only difference is $A_i$ prominences are uniform but those in $B_i$ are made of clumps.) The equational form of the initial condition is as follows,
\begin{equation}
\rho =\left\{ \begin{array}{ll}
r_{sh,cor}\rho_{cor}~~~~~(x < -2R_p) \\
\rho_{cor}  ~~~~~~~~~(-2R_p < x, r_p>R_p) \\
\rho_p  ~~~~~~~~~(-2R_p < x, r_p<R_p ) 
\end{array} \right.
\end{equation}
\begin{equation}
p =\left\{ \begin{array}{ll}
R_{sh,cor}p_0 ~~~~~(x < -2R_p) \\
p_0  ~~~~~~~~~(-2R_p < x ) 
\end{array} \right.
\end{equation}
\begin{equation}
B_x=0
\end{equation}
\begin{equation}
B_y=0
\end{equation}
\begin{equation}
B_z =\left\{ \begin{array}{ll}
r_{sh,cor}B_0 ~~~~~(x < -2R_p) \\
B_0  ~~~~~~~~~(-2R_p < x ) 
\end{array} \right.
\end{equation}
\begin{equation}
V_x =\left\{ \begin{array}{ll}
V_{sh,cor} ~~~~~(x < -2R_p) \\
0  ~~~~~~~~~(-2R_p < x ) 
\end{array} \right.
\end{equation}
\begin{equation}
V_y=0
\end{equation}
\begin{equation}
V_z=0
\end{equation}
with $\rho_{cor}=1$, $p_0=1$, $B_0=\sqrt{8\pi p_0/\beta}$. The variables $r_{sh,cor}$, $R_{sh,cor}$ and $V_{sh,cor}$ in above equations are density jump ($=$ compression ratio), pressure jump and plasma velocity of coronal shock wave, respectively. Plasma $\beta$ is assumed to be $\beta=0.2$ to model low beta corona, where Lorentz force dominates gas pressure gradient force in accelerating the prominence.
The electrical resistivity $\eta_0$ is set to be $3.0\times 10^{-4}$ in all cases, in order to prevent numerical instability.

The unit of speed in our simulation is $\sqrt{p_0/\rho_0}=1$. The corresponding sonic, Alfvenic and fast mode wave phase speeds in the corona are respectively expressed as,
\begin{equation}
C_{s,c}=\sqrt{\frac{\gamma p_0}{\rho_0}}~=~1.29
\end{equation}
\begin{equation}
C_{A,c}=\frac{B_0}{\sqrt{4\pi\rho_0}}~=~\sqrt{\frac{2}{\gamma\beta}}~=~3.16
\end{equation}
\begin{equation}
C_{f,c}=\sqrt{C_{s,c}^2+C_{A,c}^2}~=~\sqrt{1+\frac{2}{\gamma\beta}}~=~3.42
\end{equation}

From MHD Rankine-Hugoniot relations for perpendicular shocks, sonic and fast mode Mach numbers $M_{s,c}$ and $M_{f,c}$ are respectively expressed by $r_{sh,cor}, \gamma, \beta$ as,
\begin{equation}
M_{s,c}=\sqrt{\frac{2r_{sh,cor}\left(\left(2-\gamma\right)r_{sh,cor}+\gamma r_{sh,cor}\left(\beta+1\right)\right)}{\beta\gamma\left(\left(\gamma+1\right)-r_{sh,cor}\left(\gamma-1\right)\right)}}~=~3.39
\end{equation}
\begin{equation}
M_{f,c}=\sqrt{\frac{2r_{sh,cor}\left(\left(2-\gamma\right)r_{sh,cor}+\gamma\left(\beta+1\right)\right)}{\left(\left(\gamma+1\right)-r_{sh,cor}\left(\gamma-1\right)\right)\left(\gamma\beta+2\right)}}~=~1.28.
\end{equation}
From MHD Rankine-Hugoniot relations, the pressure jump $R_{sh,cor}$ and the plasma velocity of shocked corona $V_{sh,cor}$ are expressed respectively as follows.
\begin{equation}
R_{sh,cor}=\gamma M_{s,c}^2\left(1-\frac{1}{r_{sh,cor}}\right)-\frac{r_{sh,cor}^2-1}{\beta}+1~=~1.80
\end{equation}
\begin{equation}
V_{sh,cor}=(1-1/r_{sh,cor})M_{f,c}C_{f,c}~=~1.18.
\end{equation}

The simulation box is $x \in [-60 , 60]$, $y \in [0, 60]$, $z \in [0, 60]$ discretized with non-uniformly arranged $N_x\times N_y\times N_z=800\times 400\times 400$ grid points. Especially, inner region of $x \in [-7.5 , 7.5]$, $y \in [0, 4.5]$, $z \in [0, 4.5]$ is discretized with uniformly set $n_x\times n_y\times n_z=400\times200\times200$ grid points. We applied reflective boundary conditions on $y=0$ and $z=0$ planes, while for other boundaries we applied free boundary conditions. We use sparse grids in outer space so that we can neglect unwanted numerical effects on prominence dynamics from outer free boundaries.

\subsection{Momentum transport from coronal shock wave to a prominence}
Figure 4 (a) and (c) show the density distribution of two different times ($t=0$ and $t=10$, respectively) in simulation Run $A_3$. The prominence hit by the coronal fast mode shock is compressed and start to move in the direction of shock propagation (positive $x$ direction).
The fast mode shock front transmitted into the prominence material experiences multiple reflections at the prominence-corona boundary. Figure 5 (c) shows the distribution of z-component of vorticity $(\nabla\times {\bf{V}})_z$ in XY plane. In figure 5 (c), we find sharp velocity shear at the prominence-corona boundary. The velocity shear result in the development of Kelvin-Helmholtz instability in XY plane (Figure 5 (a) and (f)). The magnetic tension force suppress the the velocity shear in XZ plane (Figure 5 (d)). Associated with Kelvin-Helmholtz vortex behind the cloud, the low pressure region is formed behind the cloud, helping the prominence being accelerated in x-direction (Figure 5 (e) and (f)). Magnetic tension force also help stretch the cloud in x-direction. In Figure 4, time evolution of density structure in simulation Run $A_3$ and $B_3$ is shown with color contour in logarithmic scale together with magnetic field vectors in the plane of the plots shown by white arrows. The cloud in Run $A_3$ is deformed by Kelvin-Helmholtz instability that have developed behind the cloud. In Run $B_3$, time needed for each clump to be deformed by shear flows is much shorter thatn in Run $A_3$ because of small length scale of each clumps. In Runs $B_i$, each shocked clumps interact with each other through flow field around them making the overall flow and density structure more complicated compared with Runs $A_i$ in a short time scale.

Figure 7 shows the time evolution of the center of mass velocity $V_p$ of the prominence in Runs $A_i$ and $B_i$. The center of mass velocity of prominence $V_p$ is defined as follows,
\begin{equation}
V_p = \frac{\int_{\rho > \rho_{threshold}} \rho V_x~dxdydz}{\int_{\rho > \rho_{threshold}} \rho~dxdydz}
\end{equation}
, where $\rho_{threshold}$ is the threshold mass density and set to be 2 in all simulation runs. We regard the plasma with density $\rho >\rho_{threshold}$ as prominence in this analysis. The volume integral is done over the region where the mass density is larger than $\rho_{threshold}$.
Practically, we first flag the region where $\rho$ is larger than $\rho_{threshold}$ in the whole computational box, and then sum up the quantities within the flagged region. We tried various values of $\rho_{threshold}$, and found no significant change in the analysis results. It takes more time in Runs $A_1$ and $A_2$ for $V_p$ to approach $V_{sh,cor}$ than in Runs $B_1$ and $B_2$, although the time evolution of $V_p$ in Runs $A_3$ and $A_4$ is very similar to that in Runs $B_3$ and $B_4$.

\subsection{Phenomenological model of prominence activation}
Here, we focus on the mechanism of the momentum transfer from the coronal shock wave to the prominence. \citet{taka2015} discussed the momentum transfer mechanism of shock injection into the prominence, and estimated the resultant prominence velocity $V_{p,ss}$ with the help of one-dimensional linear theory as,
\begin{equation}
V_{p,ss} = \frac{2}{1+\sqrt{\chi}}V_{sh,cor}.
\end{equation}
, where $\chi=\rho_p/\rho_{cor}$ is the ratio of the mass density between corona and prominence.

Multidimensionality and non-linearity effects on prominence activation that are neglected in 1D model are important especially in quantitative discussion. In this section, we make a phenomenological model of prominence activation taking into consideration the effect of fluid drag force and magnetic tension force, and compare it with the result of 3D MHD simulation of prominence activation (Runs $A_i$). The schematic figure of the phenomenological model and corresponding simulation snap shots are shown in Figure 9. 
In our case, the center of mass of the prominence moves parallel to the direction of shock propagation (x-direction) because the coronal shock that activate the spherical prominences is a perpendicular one. In this section, we focus on how prominence center of mass is accelerated in x-direction as a result of the interaction with coronal perpendicular shock wave.
We define the timescale in which the injected shock front sweeps the prominence material as 'shock sweeping' timescale, $\tau_{ss}\approx 2R_p/C_{f,p}$, where $C_{f,p}=C_{f,c}/\sqrt{\chi}$ is the fast mode phase speed in the prominence. This is similar to 'cloud crushing' timescale, $\tau_{cc}$, discussed in \citet{klei1994} during which a molecular cloud is crushed by strong shock wave in interstellar medium. This shock sweeping mechanism will accelerate the prominence to have the velocity of $V_{p,ss}$ according to 1D linear theory. During the shock sweeping process ($t < \tau_{ss}$), the mean acceleration of the prominence through shock sweeping mechanism $\alpha_{ss}$ is approximated as,
\begin{equation}
\alpha_{ss}\approx A_{ss}\frac{V_{p,ss}}{\tau_{ss}}\approx A_{ss}\frac{V_{sh,cor}C_{f,c}}{\sqrt{\chi}(1+\sqrt{\chi})R_p}.
\end{equation}
$A_{ss}$ is a ad-hoc factor of order unity introduced to take into account multidimensional effect such as shock refraction and set to be $A_{ss}=0.5$ throughout the paper.
The difference between the velocity of shocked coronal plasna $V_{sh,cor}$ and the prominence itself $V_p(t)$ causes the pressure difference between the front and back side of the prominence, which accelerate the prominence to the direction of shock propagation(fluid drag force). The prominence acceleration by the fluid drag force $\alpha_{drag}(t)$ is approximated as,
\begin{equation}
\alpha_{drag}(t) \approx -\frac{C_dr_{sh,cor}\rho_{cor}S_p}{M_p}(V_p(t)-V_{sh,cor})|V_p(t)-V_{sh,cor}|
\end{equation}
, where $S_p\approx \pi R_p^2$ and $M_p\approx \frac{4\pi}{3}\rho_pR_p^3$ are cross sectional area and total mass of the prominence assuming spherical shape of the prominence, respectively. $C_d$ is a drag coefficient. In the following discussion, we approximate the drag coefficient by unity, i.e. $C_d\simeq 1$. It is revealed in the following discussion that the drag acceleration $\alpha_{drag}$ is much smaller than the other two acceleration mechanisms in weak shock acceleration, which is the case in prominence activation by coronal shocks. Substituting these into equation (24), we get,
\begin{equation}
\alpha_{drag}(t) \approx -\frac{3r_{sh,cor}}{4\chi R_p}\tilde{V_p}(t)|\tilde{V_p}(t)|
\end{equation}
, where $\tilde{V_p}(t)=V_p(t)-V_{sh,cor}$ is the prominence center of mass speed relative to ambient shocked corona.
Also, velocity difference between the prominence and ambient corona distort the magnetic field lines penetrating the prominence. This curved magnetic field accelerate the prominence by magnetic tension force. The equation of motion of the prominence accelerated by the x-component of magnetic tension force is approximately written as,
\begin{equation}
M_p\alpha_{tension}(t) \approx 2\frac{(r_{sh,cor}B_0)^2}{4\pi}S_p'(t)\tan\theta
\end{equation}
with $\alpha_{tension}$ being the acceleration by magnetic tension mechanism. Here, $S_p'(t)$ is the effective cross sectional area of the part of the prominence in XY-plane through which shocked coronal magnetic field lines penetrate. $\tan\theta = B_x/B_z$ is the inclination of distorted magnetic field lines (Figure 9). 
Assuming the inclination is determined by the balance between displacement of magnetic field lines by velocity difference $-\tilde{V_p}(t)\delta t$ and its relaxation by Alfven wave propagation $C_A\delta t$, $\tan\theta$ is approximated as,
\begin{equation}
\tan\theta \approx -\frac{\tilde{V_p}(t)}{C_A}
\end{equation}
, where $C_A\approx \frac{r_{sh,cor}B_0}{\sqrt{4\pi r_{sh,cor}\rho_{cor}}}$ is the Alfven wave phase speed in the shocked coronal plasma.
Here, we approximate $S_p'(t)$ as follows,
\begin{equation}
S_p'(t) \approx \left\{ \begin{array}{ll}
S_p\frac{t}{\tau_{sp}} ~~~~~(0< t < \tau_{sp}) \\
S_p  ~~~~~~~~~( \tau_{sp} < t ) 
\end{array} \right.
\end{equation}
, where shock passage timescale $\tau_{sp} \simeq 2R_p/C_{f,c} = \tau_{ss}/\sqrt{\chi}$ is the timescale in which the shock front pass over the prominence.
Then $\alpha_{tension}$ is finally approximated as,
\begin{equation}
\alpha_{tension}(t) \approx \left\{ \begin{array}{ll}
-\frac{3r_{sh,cor}^{3/2}\tilde{V_p}(t)C_{A,c}}{2\chi R_p}\frac{t}{\tau_{sp}} ~~~~~(0< t < \tau_{sp}) \\
-\frac{3r_{sh,cor}^{3/2}\tilde{V_p}(t)C_{A,c}}{2\chi R_p} ~~~~~~~~~( \tau_{sp} < t ) 
\end{array} \right.
\end{equation}

The order of magnitude ratio of maximum accelerations for each mechanisms is roughly $\alpha_{ss}:\alpha_{tension}:\alpha_{drag} \sim 1: 1 : (1-1/r_{sh,cor})M_{f,c}$, if $\beta$ is much smaller than unity and the shock is not strong. When the coronal shock is weak, $\alpha_{drag}$ is negligible compared with other two. After the shock front have swept the prominence, the main acceleration mechanism is magnetic tension force acceleration.

Summarizing above discussion, the prominence acceleration in the phenomenological model $\alpha_{ph}(t)$ is written as follows,
\begin{equation}
\alpha_{ph}(t) \approx \left\{ \begin{array}{ll}
\alpha_{ss}+\alpha_{drag}(t)+\alpha_{tension}(t) ~~~~~(0< t < \tau_{ss}) \\
\alpha_{drag}(t)+\alpha_{tension}(t)  ~~~~~~~~~( \tau_{ss} < t ) 
\end{array} \right.
\end{equation}
, or explicitly, $\alpha(t)$ is expressed as follows,
\begin{equation}
\alpha_{ph}(t) \approx \left\{ \begin{array}{ll}
A_{ss}\frac{V_{sh,cor}C_{f,c}}{\sqrt{\chi}(1+\sqrt{\chi})R_p}-\frac{3r_{sh,cor}\tilde{V_p}(t)|\tilde{V_p}(t)|}{4\chi R_p}+\frac{3r_{sh,cor}^{3/2}\tilde{V_p}(t)C_{A,c}}{2\chi R_p}\frac{t}{\tau_{sp}} ~~(0< t < \tau_{sp}) \\
A_{ss}\frac{V_{sh,cor}C_{f,c}}{\sqrt{\chi}(1+\sqrt{\chi})R_p}-\frac{3r_{sh,cor}\tilde{V_p}(t)|\tilde{V_p}(t)|}{4\chi R_p}+\frac{3r_{sh,cor}^{3/2}\tilde{V_p}(t)C_{A,c}}{2\chi R_p} ~~(\tau_{sp} < t < \tau_{ss}) \\
-\frac{3r_{sh,cor}\tilde{V_p}(t)|\tilde{V_p}(t)|}{4\chi R_p}+\frac{3r_{sh,cor}^{3/2}\tilde{V_p}(t)C_{A,c}}{2\chi R_p}  ~~~~~~( \tau_{ss} < t ) 
\end{array} \right.
\end{equation}

The time evolution of prominence velocity in the phenomenological model $V_{ph}(t)$ is obtained by solving the prominence equation of motion,
\begin{equation}
\frac{d V_{ph}(t)}{dt}=\alpha_{ph}(t).
\end{equation}
Substituting the simulation parameters for Run $A_i$ into the above equation of motion, we get the resultant time evolution of phenomenological prominence velocity. Time evolution of prominence velocity in Runs $A_i$ and Models $A_i$ are compared in Figure 8. 

On the other hand, the fluid equation of motion is written as,
\begin{equation}
\rho\frac{d V_x}{dt}=-{\partial_x}(p+\frac{B^2}{8\pi})-\frac{({\bf B}\cdot\nabla)B_x}{4\pi}
\end{equation}

If we assume the prominence mass $M_p=\int_{\rho > \rho_{threshold}} \rho_p dxdydz$ is constant, the volume integral of the above equation of motion over the prominence volume leads to the following relations,
\begin{equation}
\alpha(t)=\alpha_{tp}(t)+\alpha_{mt}(t)
\end{equation}
, with $\alpha(t)$, $\alpha_{tp}(t)$ and $\alpha_{mt}(t)$ being  the acceleration of center of mass of the prominence in x-direction, prominence acceleration by total pressure gradient force and prominence acceleration of magnetic tension force, respectively. $\alpha(t)$, $\alpha_{tp}(t)$ and $\alpha_{mt}(t)$ are respectively written (and calculated from simulation results) as follows,
\begin{equation}
\alpha(t)=\frac{1}{M_p}\int_{\rho > \rho_{threshold}}\rho \frac{dV_x}{dt}{}dxdydz
\end{equation}
\begin{equation}
\alpha_{tp}(t)=-\frac{1}{M_p}\int_{\rho > \rho_{threshold}}\partial_x(p+\frac{B^2}{8\pi})dxdydz
\end{equation}
\begin{equation}
\alpha_{mt}(t)=-\frac{1}{M_p}\int_{\rho > \rho_{threshold}}\frac{({\bf B}\cdot\nabla)B_x}{4\pi}dxdydz
\end{equation}

In Figure 10 (a)-(h), we show the total pressure gradient acceleration $\alpha_{tp}$ and magnetic tension acceleration $\alpha_{mt}$ with dashed and solid lines, respectively in MHD simulation Runs $A_i$ and $B_i$. Figure 10 (i)-(l) show $\alpha_{ss}+\alpha_{drag}(t)$ and $\alpha_{tension}$ in Models $A_i$. Sudden drop in total pressure gradient force acceleration in for example Run $A_2$ (Figure 10 (c)) corresponds to the first reflection of the wave which have swept the whole prominence (Figure 9 (e)). The oscillations of accelerations in Runs $A_i$ are the results of multiple wave reflections at prominence-corona boundary (Figure 9 (f)), and the oscillation period is approximately $\tau_{ss}$ for each Runs. In Run $B_1$ and $B_2$, the total pressure gradient force acceleration is smaller than that in Run $A_1$ and $A_2$. That is one reason why it took more time for the prominence velocity $V_p$ in Runs $A_1$ and $A_2$ to approach the velocity of shocked corona $V_{sh,cor}$ than in Runs $B_1$ and $B_2$. After the shock sweeping time $\tau_{ss}$, $\alpha_{ss}+\alpha_{drag}(t)$ in the phenomenological models $A_i$ are smaller than pressure gradient force acceleration $\alpha_{tp}$ in Runs $A_i$. Also, the magnetic tension force acceleration $\alpha_{tension}(t)$ in the phenomenological models $A_3$ and $A_4$ underestimates the actual simulation results $\alpha_{mt}$ in Runs $A_3$ and $A_4$. Roughly, the time evolution of $\alpha_{tp}$ and $\alpha_{mt}$ in Runs $A_i$ resemble to that of $\alpha_{ss}+\alpha_{drag}(t)$ and $\alpha_{tension}$ in Models $A_i$, respectively. This is natural because $\alpha_{ss}$ in Models $A_i$ represents the acceleration by fast mode shock transmission into the prominence, which is mainly due to total pressure gradient. Also, $\alpha_{drag}$ in Models $A_i$ is mainly due to pressure gradient which lasts longer than $\tau_{ss}$ resulting from velocity difference between corona and prominence. Due to velocity shear at the prominence-corona boundary in XY plane, KHI develops and low pressure region is formed behind the cloud associated with induced vortices (Figure 5 (a), (c) and (e)). Also, we see a high pressure region ahead of the cloud due to flow collision. This type of flow structure result in the large scale total pressure gradient that accelerate the prominence in x-direction as a drag force. Formation of induced vortices and associated low density region behind the cloud is clearer in hydrodynamic case (simulation Run $C_3$) discussed in section 3.6 (Figure 13).

\subsection{Excitation and damping of internal flow}
While coronal shock wave gives its bulk momentum to the prominence, internal flow is also excited within the prominence. Internal flow excitation during interstellar shock-cloud interaction has been studied widely as a measure of internal turbulence. The internal flow is discussed using velocity dispersion in x, y and z components, 
\begin{equation}
\delta V_x = \sqrt{\left<V_x^2\right> - \left<V_x\right>^2}
\end{equation}
\begin{equation}
\delta V_y = \sqrt{\left<V_y^2\right>}
\end{equation}
\begin{equation}
\delta V_z = \sqrt{\left<V_z^2\right>}
\end{equation}
, where mass weighted average of physical quantity $f$ is calculated as,
\begin{equation}
\left<f\right> = \frac{\int_{\rho > \rho_{threshold}} \rho f~dxdydz}{\int_{\rho > \rho_{threshold}} \rho~dxdydz}
\end{equation}
We assumed $\left<V_y\right>$ and $\left<V_z\right>$ should be $0$ because of the symmetry around x-axis.

Figure 11 shows the time evolution of velocity dispersion $\delta V_x$, $\delta V_y$ and $\delta V_z$ in Runs $A_i$ and $B_i$ in semi-log graph. They show the excitation of the internal flow through the passage of shock, and its damping roughly in an exponential manner. In Runs $A_i$, we also find oscillatory behavior of $\delta V_x$ with their period of roughly $0.5\tau_{ss}$. We estimated the damping times (e-folding times) $\tau_d$ for $\delta V_x$, $\delta V_y$ and $\delta V_z$ in Runs $A_i$ and $B_i$ based on least square fitting by bisector method \citep{iso1990} between $t=10$ and $t=20$. The resultant slopes for exponential damping are shown as thick lines for each variables in Figure 11. We did not estimate them in Runs $A_4$ and $B_4$ because the simulation time span did not cover their damping phase. The estimated damping times are shown in Table 1. The estimated damping times fall in the range between $\tau_{ss}$ and 2$\tau_{ss}$ in almost all the cases. $\tau_{d,vz}$ is roughly twice as large as $\tau_{d,vy}$ and $\tau_{d,vx}$. 

We show the time evolution of the center of mass position of the prominence $<x>$ in Figure 12. Uniform mesh is used between $X=\pm 7.5$ shown as dashed lines in Figure 12. In all eight simulation Runs, the prominence acceleration phase is mostly solved within uniform mesh region. In Figure 12, we have no significant change of damping behavior when the cloud center of mass crossed $X=7.5$ plane.

\subsection{Comparison with hydrodynamic simulation}
In this section, we study hydrodynamic simulation of shock-cloud interaction (Run $C_3$) and compare it with MHD simulation Run $A_3$. By directly comparing Run $C_3$ with Run $A_3$, we discuss hydrodynamic effects which were suppressed in MHD Run $A_3$ by the presence of magnetic field. Such hydrodynamic effects are expected to be more significant in interstellar shock-cloud interaction.
In practice, we modified the initial plasma beta in Run $A_3$ to $1.0\times 10^{10}$ and use it as an initial condition of the MHD simulation. The pressure jump $R_{sh,cor}$, plasma velocity of shocked corona $V_{sh,cor}$ and shock Mach number $M_{s,c}$ are $R_{sh,cor}=1.70$, $V_{sh,cor}=0.44$ and $M_{s,c}=1.25$, respectively. 
The minimum plasma beta appeared throughout the simulation was $2.7\times 10^4$ which is still much larger than unity. The magnitude of plasma beta is of the order of the ratio of pressure gradient force term to Lorentz force term in MHD equation of motion. Because plasma beta is very large ($>10^4$) throughout the simulation, the effect of magnetic field on the dynamics in simulation Run $C_3$ is negligible. So, we call Run $C_3$ ``hydrodynamic'' simulation further on.

Figure 13 (a) shows the volume rendering of the density distribution in Run $C_3$ at $t=60$($=2.3\tau_{ss}$). 
After being crushed by the injected shock wave, the cloud in Run $C_3$ continues to be deformed.
This is quite different from the time evolution of cloud shape in MHD Run $A_3$. In Run $A_3$, magnetic field suppress the cloud deformation and damps the internal flow rapidly in the time scale of $\sim\tau_{ss}$. Without such a magnetic suppression of cloud deformation, the time evolution of the cloud as well as ambient flow structure in Run $C_3$ is very different with those in Run $A_3$. One characteristic flow structure in Run $C_3$ is the formation of large coherent vortex flow behind the cloud. The ambient flow converging towards the cloud along x-axis and diverging in YZ plane associated with the vortex helps flatten and stretch the cloud further in Run $C_3$ (Figure 13 (b) and (c)).

In Figure 14 (a)-(d), snapshots of density and pressure distribution in XZ plane at $t=0.31 \tau_{ss}$ in simulation Run $A_3$ and $C_3$ are shown, respectively. The velocity amplitudes of transmitted shock front $V_{sf1}$ and that of the shock front from behind the cloud $V_{sf2}$ in simulation Run $A_3$ and $C_3$ are shown in Figure 14 (e) and (f), respectively. The shock front 2 appears when the injected shock wave have passed over the cloud at $t\approx \tau_{sp}$, where $\tau_{sp}=2{R_p}/C_{f,c}$ is the shock passage timescale. The ratio of $V_{sf2}$ to $V_{sf1}$ in MHD cases are much smaller than unity. For example, $V_{sf2}/V_{sf1}$ is 0.29 at $t=3$($=0.31\tau_{ss}=1.7\tau_{sp}$) in Run $A_3$ (Figure 14 (e)). In hydrodynamic case, on the other hand, $V_{sf2}$ is similar to $V_{sf1}$. $V_{sf2}/V_{sf1}$ is 0.70 at $t=8$($=0.31\tau_{ss}=1.7\tau_{sp}$) in Run $C_3$ (Figure 14 (f)). In hydrodynamic case, shock sweeping acceleration mechanism that is driven by the propagation of shock front 1 in Figure 14 (f) is almost canceled out by propagation of shock front 2 after $t=\tau_{sp}$.

Taking above effect into account, the phenomenological model of cloud acceleration in hydrodynamic case is modified from Model $A_i$ to be as follows,
\begin{equation}
\alpha_{ph}(t) \approx \left\{ \begin{array}{ll}
A_{ss}\frac{V_{sh,cor}C_{s,c}}{2\sqrt{\chi}(1+\sqrt{\chi})R_p}+\frac{3r_{sh,cor}(V_{sh,cor}-V_p(t))^2}{4\chi R_p}~~~~~(0< t < \tau_{sp}) \\
\frac{3r_{sh,cor}(V_{sh,cor}-V_p(t))^2}{4\chi R_p}  ~~~~~~~~~( \tau_{sp} < t ) 
\end{array} \right.
\end{equation}

The time evolution of prominence velocity $V_p$ in simulation Run $C_3$ is shown as a solid line of Figure 15 (a), together with that of the phenomenological model Model $C_3$ discussed above shown as a dashed line. Model $C_3$ captures abrupt acceleration characteristics before $t = \tau_{sp}$, but underestimates the acceleration in the later phase. In Model $C_3$, the acceleration decrease because of relative velocity $V_{sh,cor}-V_p$ become smaller with time, but in Run $C_3$, the acceleration is almost constant after $t = \tau_{sp}$. This is partly because of the deformation (flattening and streaching) of the prominence in hydrodynamics case. If we denote the cross sectional area of the cloud in YZ plane as $S_{yz}(t)$, the magnitude of aerodynamic drag force that mainly accelerate the cloud in Run $C_3$ is of the order of $\sim \rho_{sh,cor}\tilde{V_p}(t)^2S_{yz}(t)$.
In Model $C_3$, we assumed the prominence has a spherical shape, although in Run $C_3$, the cloud is flattened and stretched in YZ plane. We study $<r^2>=<y^2+z^2>$ as an effective cross sectional area of the prominence in YZ-plane where aerodynamic drag works to accelerate the prominence in x-direction. Figure 15 (c) shows the time evolution of $<r^2>$. $<r^2>$ at $t=90$ is almost 5 times that of initial $<r^2>$. We note that in Run $C_3$, drag acceleration mechanism works much more than in Model $C_3$ because $S_{yz}\sim <r^2>$ evolves in time as discussed above (Figure 15 (c)).

The internal flow structure is also different in hydrodynamic case compared with MHD case. Figure 15 (d) shows time evolution of velocity dispersions $\delta V_x =\sqrt{<V_x^2> - <V_x>^2}$ and $\delta V_y =\sqrt{<V_y^2>}$ as a proxy of internal flow within the prominence. In Run $C_3$, $\delta V_x$ and $\delta V_y$ increase with time after the shock passage. This is different from the results in Run $A_3$ where the velocity dispersions damp in an exponential manner due to the presence of magnetic field.

\subsection{Prominence activation by triangular wave packet}
Coronal shock waves that activate prominences in reality are not blast waves but wave packets with finite width. The motion of activated prominence depends not only on plasma velocity of the shock but also on wave packet width. In this secsion, we analyze 3D MHD simulation results that reproduced prominence activation by triangular wave packet, and compare with phenomenological model. 
The density profile of the prominence in this simulation (let's call it Run $D_3$ further on) is the same with that in simulation Run $A_3$. In the simulation Run $D_3$, we have a triangular wave packet with velocity amplitude $V_{sh,cor}$ and wave packet width $w$ that activate prominence.
The initial plasma velocity distribution for simulation Run $D_3$ is as follows (solid line in Figure 16 (a)),
\begin{equation}
V_x =\left\{ \begin{array}{ll}
2V_{sh,cor}\frac{x+(2R_p+2w)}{2w} ~~~~~(-2R_p-2w<x<-2R_p) \\
0  ~~~~~~~~~~~~~(x<-2R_p-2w,-2R_p< x) 
\end{array} \right.
\end{equation}
, with the wave packet width being $w=10$ (Figure 16 (a)). The density, pressure and magnetic field in the corona are all uniform, initially (solid lines in Figure 16 (b), (c) and (d)).

The initial condition is a ``superposition'' of two triangular fast mode wave packets propagating in opposite directions to each other with velocity amplitude $V_{sh,cor}$ and wave packet width $w$. If the wave packets were linear ones (i.e. $v_{sh,cor}<<C_{f,c}$), the wave propagating in the positive x-direction would keep its velocity amplitude and wave packet width unchanged without any interaction with oppositely-directed wave packet. In simulation Run $D_3$, the wave packet propagating in the positive x-direction interact with the prominence.

We checked by nonlinear 1D MHD numerical simulation (without prominence) how initially superposed wave packets (shown as solid lines in Figure 16) evolve in time. Dashed and dash-dotted lines in Figure 16 denote plasma parameters distribution at times $t=2$ and $t=6$, respectively in the 1D simulation (Figure 16). We see in Figure 16 (a) that the initial single peak of superposed wave packets in plasma velocity split into two oppositely-directed wave packets as expected from linear theory. Because of the non-linearity, however, the velocity amplitude decays slowly and the wave packet broadens as they propagate.

We make the phenomenological model that describes prominence center of mass motion in simulation Run $D_3$. We call it Model $D_3$. The Model $D_3$ can be obtained simply by replacing $V_{sh,cor}$ and $\tilde{V_p}(t)=V_p(t)-V_{sh,cor}$ in Model $A_3$ with $V_{wp,cor}(t)$ and $\tilde{V_p}(t)=V_p(t)-V_{wp,cor}(t)$, respectively. $V_{wp,cor}(t)$ is the plasma velocity in the corona around the prominence shocked by triangular wave packet and approximated as follows,
\begin{equation}
V_{wp,cor}(t) =\left\{ \begin{array}{ll}
V_{sh,cor}\frac{2\tau_{wp}-t}{2\tau_{wp}} ~~~~~(0<t<2\tau_{wp}) \\
0  ~~~~~~~~~~~~~~~~~~~~~~~~~(2\tau_{wp}< t) 
\end{array} \right.
\end{equation}
with $\tau_{wp}=w/C_{f,c}$ being the ``wave packet passage'' time scale. $V_{wp}$ is an approximation based on the fact that the injected coronal shock is weak and the activated prominence moves much slower than coronal fast mode phase speed.
The solid and dashed lines in Figure 17 (a) show prominence center of mass acceleration by magnetic tension force $\alpha_{mt}$ and that by total pressure gradient force $\alpha_{tp}$ in simulation Run $D_3$, respectively. The solid and dashed lines Figure 17 (b) show prominence center of mass acceleration by magnetic tension mechanism $\alpha_{tension}$ and that both by shock sweeping and fluid drag mechanisms $\alpha_{ss}+\alpha_{drag}$, respectively. We find from the plot in Figure 17 (a) that the prominence is mainly accelerated by magnetic tension force first, and then decelerated also by magnetic tension force. The Model $D_3$ captures such a characteristic response of the prominence to wave packet injection. We find some oscillations in both $\alpha_{mt}$ and $\alpha_{tp}$ in simulation Run $D_3$. They result from multiple reflections of wave packets within the prominence. The effect of multiple reflections is not included in Model $D_3$. 

Solid and dashed lines in Figure 18 (a) show time evolution of prominence center of mass position $X_p$ in simulation Run $D_3$, and that expected from model $D_3$, respectively. Solid and dashed lines in Figure 18 (b) shows the prominence center of mass speed $V_p$ in simulation Run $D_3$, and that expected from model $D_3$, respectively. From Figure 18, we see that the center of mass motion of activated prominence expected with Model $D_3$ quantitatively agreed with those in simulation run $D_3$. We call the time interval during which the prominence is accelerated to its maximum speed by the shock wave as ``acceleration phase'' of the prominence activation. The acceleration phase continued until $t\simeq4$ in Run $D_3$ (Figure 18 (b)). Solid, dashed and dash-dotted lines in Figure 19 show time evolution of velocity dispersions $\delta V_x$, $\delta V_y$ and $\delta V_z$ in simulation Run $D_3$, respectively. Exponential damping timescales are estimated for the three velocity dispersions by least square fit by bisector method during the time between $t=10$ and $t=20$. The resultant slopes for exponential damping are shown as thick lines for each components. Damping times $\tau_{d,vx}$, $\tau_{d,vy}$ and $\tau_{d,vz}$ are listed in the last row in table 1. We find that the damping times for all three components are similar to the shock sweeping time scale $\tau_{ss}$. We note that $\tau_{d,vz}$ in simulation Run $D_3$ is much smaller than that in Run $A_3$.

\section{Coronal shock and prominence diagnostics using prominence activation}
In this section, we try to diagnose coronal shock properties and prominence properties using prominence activation with the help of phenomenological model discussed above. The Figure 20 (a) is a time-distance diagram of prominence activation made from AIA 193$~{\AA}$ pass band images. This corresponds to the white rectangle in Figure 3 (i). The white crosses denote prominence positions during the prominence activation estimated by the eye. We denote the prominence displacement in the plane of the sky at time $t=t_i$ ($i=1-14$) as $L_{p,i}$. On the other hand, the position of the prominence activated by triangular wave packet can be predicted by the phenomenological model discussed in the previous section. 

Here, we try to fit observed time evolution of activated prominence position with the phenomenological model expectation. 
I used equation 31 modified with $V_{wp,cor}$ as described in section 3.7 for the fitting.
We assume coronal temperature to be $T=10^6$ K. This leads to the coronal sound speed to be $C_{s,c}\simeq 1.8\times 10^2$ km s $^{-1}$, assuming the specific heat ratio to be $\gamma=5/3$. The local density gap between corona and prominence is assumed to be $\chi=100$. We think of two different values for the angle $\phi$ between the line of sight and shock propagation direction, that are $\phi=0^{\circ}$, $45^{\circ}$. The thickness of the prominence core seen as dark structure in AIA 193 $~{\AA}$ images are about 10'', which correspond to the estimated prominence radius of $R_p\simeq 3.6\times10^3$ km. The fast mode wave speed in the corona $C_{f,c}$ is estimated to be $C_{f,c}\simeq C_{wv}/\cos\phi$, with $C_{wv}=380$ km s$^{-1}$ being wave propagation speed in the plane of sky. The coronal fast mode wave speeds in $\phi=0^{\circ}$ and $45^{\circ}$ cases are $3.8\times10^2$ km s$^{-1}$ and $5.4\times10^4$ km s$^{-1}$, respectively. The plasma beta is obtained with $\beta=(2/\gamma)(C_{s,c}/C_{A,c})^2$, with $C_{A,c}=\sqrt{C_{f,c}^2-C_{s,c}^2}$ is a coronal Alfven speed in perpendicular propagation case. In $\phi=0^{\circ}$ and $\phi=45^{\circ}$ cases, plasma beta is calculated to be $\beta=0.33$ and $\beta=0.15$, respectively. Assuming shock propagation direction and the center of mass velocity of activated prominence is parallel (which is correct in perpendicular shock case), the displacements of the activated prominence at time $t=t_i$ are estimated to be $X_{p,i}=L_{p,i}/\cos\phi$. Then, the number of remaining free parameters of the phenomenological model are three, that are prominence volume filing factor $f_V$, compression ratio of coronal shock wave $r$ and wave packet width $w$. We denote the time evolution of activated prominence position expected from phenomenological model as $X_{p,model}(f_V,r,w;t)$. We searched best-fit values for $f_V,r$ and $w$ with which the sum of squared residuals $SSR(f_V,r,w)=\Sigma_{t_i} (X_{p,model}(f_V,r,w;t_i)-X_{p,i})^2$ is minimized in the parameter space $f_V\in[0.01,1.0]$, $r=[1.07,1.9]$ and $w=[0.02R_s,0.4R_s]$, with $R_s$ being a solar radius. The fitting results for $\phi=0^{\circ}$ and $\phi=45^{\circ}$ cases are shown in Figure 20 (b) and (c).

As a best-fit parameters, we estimate $r$ and $w$ of coronal shock wave to be 1.17 and 0.16 $R_s$ in $\phi=0^{\circ}$ case and 1.17 and $0.22 R_s$ in $\phi=45^{\circ}$ case, with $R_s$ being solar radius. Best-fit parameters for $\phi=0^{\circ}$ and $\phi=45^{\circ}$ cases are listed in Table 2. Fast mode Mach number $M_f$ of the coronal shock in $\phi=0^{\circ}$ and $\phi=45^{\circ}$ cases are 1.12 and 1.13, respectively. From mass conservation at the shock front, velocity amplitude of injected triangular wave is expressed as $V_{sh,cor}=M_{f,c}C_{f,c}(1-1/r)$. From this, the plasma velocity amplitude for injected triangular wave is estimated as 62 km s$^{-1}$ and 88 km s$^{-1}$ in $\phi=0^{\circ}$ and $\phi=45^{\circ}$ cases, respectively. The best-fit $f_V$ in both cases were $0.01$, which is the smallest value in the free parameter space. 

Based on the phenological model (equation 31), on the other hand, when $f_V$ is much smaller than unity, the prominence is accelerated to its maximum speed almost within the shock sweeping time scale $\tau_{ss}$, while the subsequent prominence deceleration occurs within the wave packet passage timescale $\tau_{wp}$. With the estimated parameters of $f_V=0.01$, $R_p=3.6\times10^{3}$ km, $C_{f,c}\simeq500$ km s$^{-1}$ and $w=0.2R_s$, the acceleration and deceleration timescales are roughly, $\tau_{ss}\sim20$s and $\tau_{wp}\sim300$s, respectively. 
In order to estimate $f_V$ correctly, we have to time-resolve the acceleration phase whose time scale $\sim\tau_{ss}$ reflects $f_V$ directly. Compared with the estimated acceleration timescale of $\sim20$ s, the AIA time cadence of $12$ s is not high enough to track the acceleration phase of the prominence activation in this event, although we can track the subsequent deceleration phase with sufficient time resolution. As seen in the time-distance plot of Figure 20 (a), the prominence appears to be accelerated to its maximum speed right after the arrival of the shock. We think that the evaluated value of $f_V$ in this analysis is not accurate enough due to the lack of fully time-resolved observation of the acceleration phase of the prominence activation in this event.

Then, we estimate the energy of the coronal shock wave $E_{sh,cor}$ associated with the X5.4 flare. The coronal emission measure near the prominence before the arrival of the shock was $EM\simeq1\times10^{27}$ cm$^{-5}$. The emission measure is calculated at point $A$ in Figure 3 (a), based on a method proposed in \citet{cheung2015}. Assuming a line-of-sight distance $d$ to be of order of the distance to the (solar) horizon as seen from a point above the photosphere by a pressure scale hight ($h=50$Mm), we get $d\sim\sqrt{h(2R_s+h)}\sim 2\times 10^2$Mm. The coronal proton number density $n$ at point $A$ is estimated from the relation $EM\simeq n^2 d$ to be $n \sim 2\times10^8$ cm$^{-3}$. Corresponding coronal mass density is $\rho\simeq n m_p=3\times10^{-16}$ g cm$^{-3}$, with $m_p\simeq 1.7\times 10^{-24}$ g being the proton mass. The energy flux of the shock at point A is $F_{sh,cor}\sim \rho V_{sh,cor}^2C_{f,c}\sim 1\times10^6$ erg cm$^{-2}$ s$^{-1}$. The time scale for the triangular wave to pass through the fixed point A is $\tau\sim w/C_{f,c}\sim 3\times 10^2$ s. The surface area of the spherically expanding dome of shock front in the corona is approximated as $S\sim2\pi L^2$, with L being the distance between flaring AR and point A. We approximate $L\sim R_s$ and get $S\sim 3\times 10^{22}$ cm$^{2}$. From above, the energy budget of the coronal shock wave is estimated as $E_{sh,cor}\sim F_{sh,cor}\tau S\sim 10^{31}$ erg. The total energy released in X5.4 class flare is roughly estimated by empirical relation between flare energy and flare soft X-ray peak flux to be several $10^{32}$ erg \citep{emslie2012,kletz2010}. The estimated energy budget of the globally propagating shock wave in the corona is a few percent of total energy released during the flare.

\section{Summary and discussion}
Recent high time and spatial resolution EUV observation of solar corona by {\it SDO}/AIA enabled us to study in detail the time evolution of coronal shock wave associated with flares. We can now study as well the interaction between coronal shock wave and prominences using AIA. In this paper, we studied the excitation process of large amplitude prominence oscillation through the interaction between the prominence and coronal shock wave, with the help of three-dimensional MHD simulation.

The X5.4 class flare occurred on March 7, 2012 was associated with very fast CME with its speeds of about $2,700$ km s$^{-1}$ estimated from coronagraph observation by SOHO/LASCO. A global shock front is formed around the expanding CME ejecta. The shock front had a dome-like form, especially with bright structure propagating to the north at the foot of the dome. The northward disturbance hit a polar prominence, leading to the excitation of large amplitude prominence oscillation. During the prominence activation, the prominence was strongly brightened, receiving momentum in the direction of shock propagation.

In order to explain the observational signature of prominence activation, we have done a three-dimensional MHD simulation of coronal fast mode shock-prominence interaction. Especially, the momentum transfer mechanism from the shock to the prominence is studied in detail. The shock injection into the prominence material compressed and accelerated the prominence. The velocity shear at the corona-prominence boundary resulted in Kelvin-Helmholtz instability. KHI was stabilized by magnetic tension force in the plane containing the initial magnetic field lines. By analyzing the simulation results and comparing them with phenomenological models, magnetic tension force acceleration was also found to be very important. The accelerated prominence velocity asymptotes to the value of coronal shocked plasma velocity when the shock is a blast wave after some timescale depending on different prominence density. When the volume filling factor is small like in Runs $B_1$ and $B_2$, the acceleration timescale is longer compared with the case with uniformly distributed prominence density in Runs $A_1$ and $A_2$. 
This may be because of the suppression of shock sweeping acceleration mechanism in 'clumpy' cloud in Runs $B_i$. Both total surface area of clumps and local density gap between clump and corona in Runs $B_i$ are larger than those in Runs $A_i$. This make it difficult for injected shock fronts in Runs $B_i$ to penetrate deep into the cloud as a whole so that they could exchange momentum with cloud materials. When the volume filling factors are larger than 0.3, the resultant time evolution of the mean velocity in Runs $B_3$ and $B_4$ are very similar to that of Runs $A_3$ and $A_4$. 

We also studied the time evolution of the velocity dispersion of shocked prominence material in each ($x$, $y$ and $z$) component. The velocity dispersion is excited during the shock sweeps through the cloud and then damps almost exponentially. The exponential damping time scales of velocity dispersions in each component $\tau_{d,vx}$, $\tau_{d,vy}$ and $\tau_{d,vz}$ are estimated and summarized in Table 1. In almost all the simulation runs, $\tau_{d,vx}$ and $\tau_{d,vy}$ are roughly comparable to the shock-sweeping time scale $\tau_{ss}$ while $\tau_{d,vz}$ is about twice as large as $\tau_{ss}$. A possible reason for the discrepancy of the damping times among the components is as follows. When the randomized flow is directed to positive x-direction at a certain time and location, the magnetic field lines originally directed to z-direction ($B_z$) will be distorted there in XZ plane resulting in the electric current directed in negative y-direction (-$J_y$). The Lorentz force $\sim -J_y\times{}B_z$ act in negative $x$-direction which pull back the flow originally directed to positive x-direction. This helps damp the velocity dispersion in x-direction, making $\tau_{d,vx}$ small. The same damping mechanism works on the y-component of velocity dispersion as well, but not on z-component.

In interplanetary shock-cloud interaction, both shock sweeping mechanism and fluid drag force accelerate the cloud.
KHI is also important in mixing MC materials which will affect star formation process taking place. In solar coronal shock-prominence interaction, magnetic tension force is more important in accelerating the prominence than fluid drag force, because in prominence activation, plasma beta is typically smaller than unity and the shock not strong. When the prominence has internal density structure like in Runs $B_i$, plasma mixing might also occur in short time scale.

When the plasma beta is much larger than unity (which is a reasonable assumption in some molecular clouds), the cloud acceleration and internal flow excitation shows much different characteristics. The shock-sweeping acceleration mechanism is effective only during the shock passage time $\tau_{sp}$, and the pressure gradient force due to the velocity difference between the cloud and the ambient plasma works as an accelerator. The cloud is flattened with the help of ambient flow that converge towards the cloud along x-axis and diverges in YZ plane associated with a coherent vortex formed behind the cloud.
The flattening effect increases the effective cross section of the cloud, helping the cloud acceleration by fluid drag. The excited internal flow does not decay in hydrodynamic simulation Run $C_3$, though in MHD simulation Runs $A_3$, the internal flow damps in an exponential manner mainly due to the Lorentz force.

In reality, the coronal shock wave that activate a prominence is not a blast wave as studied in Runs $A_i$ or $B_i$ but a wave packet with finite wave packet width. We studied interaction between a coronal shock wave in the form of triangular-shaped wave packet and a prominence in simulation Run $D_3$. 

The shocked prominence is first accelerated and then decelerated by magnetic tension and total pressure gradient force in the simulation Run $D_3$. The phenomenological model $D_3$ well captures the characteristic dynamics of the prominence center of mass both in acceleration and deceleration phases, but slightly underestimate the impact of pressure gradient force. Especially, the phenomenological model $D_3$ does not reproduce the prominence deceleration by total pressure gradient force which is present in Run $D_3$. One of the possible reason for the discrepancy is that the phenomenological model neglects the effect of multiple reflection of transmitted shock wave within the prominence.
We compared the prominence center of mass position and velocity in simulation Run $D_3$ with those expected by phenomenological Model $D_3$ and found quantitative agreement between the simulation and the model.

We tracked the time evolution of the position of the activated prominence and fitted it with phenomenological model. The best-fit curve for prominence movement agreed well with observation. As a best fit parameters, we obtained prominence volume filling factor $f_V$, coronal shock compression ratio $r$ and wave packet width $w$. The resultant compression ratio and fast mode mach numbers of the coronal shock were 1.17 and 1.12 in $\phi=0^{\circ}$ case, and were 1.17 and 1.13 in $\phi=45^{\circ}$ case, respectively. The estimated wave packet width of the coronal shock were 0.16$R_s$ and 0.22$R_s$ in the cases of $\phi=0^{\circ}$ and $\phi=45^{\circ}$, respectively. They are comparable to typical EUV wave front widths of $\sim100$Mm which are suggestive of coronal shock waves reported in \citet{muhr2014}. Both the estimated coronal shock propagation speed and the plasma velocity amplitude of $380-540$ km s$^{-1}$ and $68-88$ km s$^{-1}$ are reasonable values as a weak fast mode coronal shock wave. The shock wave that activated the prominence had likely been driven by the lateral expansion of CME ejecta in the lower corona (whose speed is much smaller than the radial ejection speed) and had propagated a considerable distance of $\sim 1R_s$ from the source active region. This is a possible reason why the coronal shock near the prominence was weak although the associated CME was extremely fast with its speed of almost $\sim 3000$ km s$^{-1}$ at a hight of  $2R_s$ estimated with {\it SOHO}/LASCO coronagraph observations.\footnote{http://cdaw.gsfc.nasa.gov/CME\_list/UNIVERSAL/2012\_03/univ2012\_03.html}
The best-fit value of prominence filling factor on the other hand was $f_V=0.01$, which we don't think is accurate partly because the observational data did not time-resolved the acceleration phase of activated prominence, which is vital for determining prominence $f_V$ based on our model. With the help of emission measure analysis, we estimated the energy of the coronal shock to be $E_{sh,cor}\sim 10^{31}$ erg in this event. This was roughly several percent of total released energy during the X5.4 flare.

The physical mechanism that mainly work in prominence activation differs with different wave packet width. If the width of the wave packet is longer than $C_{f,c}\tau_{ss}$, magnetic tension force is the most important in accelerating the prominence. The coronal magnetic loop that support the prominence material against gravity in the corona is rooted at the photosphere on both sides, which will reduce magnetic tension force acceleration after Alfven travel time $\tau_{A}=\frac{L}{C_{A,c}}$ where L is the length of the magnetic loop. We note that $\tau_A$ is of order of period of LAPO ensuing the prominence activation.

In this paper, we discussed the dynamics of prominence-coronal shock interaction which leads to LAPO. The prominence activation event we studied was triggered by the arrival of coronal shock wave at a polar prominence that propagated from faraway AR corona. Such globally propagating flare-associated shock waves in the corona or LAPOs are relatively rare phenomena. On the other hand, small scale magnetic explosions (small flares and jets) always occur in the corona. We expect the interactions between solar prominences and small amplitude shock waves generated by such small scale magnetic explosions are always occurring in the corona, and might play a role in driving small amplitude prominence dynamics such as small amplitude oscillations and chaotic movements of plasma elements.

\acknowledgements
The SDO/AIA data are courtesy of NASA/SDO and AIA science team. The simulation code used in this work is created with the help of HPCI Strategic Program. Numerical computations were carried out on Cray XC30 at Center for Computational Astrophysics, National Astronomical Observatory of Japan. The author is gratefull to Dr. Kazunari Shibata and Dr. Ayumi Asai of Kyoto University for their helpful comments and discussions. The author is also grateful to the journal referee for his/her careful reading and lots of comments which significantly improved the paper. This work is financially supported by the Grant-in-Aid for JSPS Fellows 15J02548.

\clearpage
\clearpage

\begin{table}
\caption{Simulation parameters and Values\label{tbl-1}}
\begin{tabular}{crrrrrrrrrrr}
\tableline
\tableline
\multicolumn{1}{c}{} & \multicolumn{1}{c}{$f_V$\tablenotemark{a}} & \multicolumn{1}{c}{$~~\overline{\rho_p}$\tablenotemark{b}} &
\multicolumn{1}{c}{$~~\tau_{ss}$\tablenotemark{c}} & \multicolumn{1}{c}{$\tau_{d,vx}\tablenotemark{d}$} & \multicolumn{1}{c}{$\tau_{d,vy}$\tablenotemark{e}} & \multicolumn{1}{c}{$\tau_{d,vz}$\tablenotemark{f}} &
\multicolumn{1}{c}{$\tau_{d,vx}/\tau_{ss}$\tablenotemark{g}} & \multicolumn{1}{c}{$\tau_{d,vy}/\tau_{ss}$\tablenotemark{h}} & \multicolumn{1}{c}{$\tau_{d,vz}/\tau_{ss}$\tablenotemark{i}}\\
\tableline
Run $A_1$ &$~~$  0.05 &$~~$  5.95 &$~~$ 4.28 &$~$  4.28 &$~$ 4.57 &$~~$ 9.14 &$~~$ 1.00 &$~$ 1.07 &$~$ 2.14 &$~$ \\
Run $A_2$ &$~~$  0.10 &$~~$  10.9 &$~~$ 5.79 &$~$  8.06 &$~$ 7.19 &$~~$ 7.71 &$~~$ 1.39 &$~$ 1.24 &$~$ 1.33 &$~$ \\
Run $A_3$ &$~~$  0.30 &$~~$  30.7 &$~~$ 9.72 &$~$  12.4 &$~$ 14.8 &$~~$ 26.6 &$~~$ 1.28 &$~$ 1.52 &$~$ 2.73 &$~$ \\
Run $A_4$ &$~~$  1.00 &$~~$  100. &$~~$ 17.5 &$~$  --- &$~$ --- &$~~$ --- &$~~$ --- &$~$ --- &$~$ --- &$~$ \\
Run $B_1$ &$~~$  0.05 &$~~$  5.95 &$~~$ 4.28 &$~$  5.31 &$~$ 3.30 &$~~$ 8.67 &$~~$ 1.24 &$~$ 0.77 &$~$ 2.03 &$~$ \\
Run $B_2$ &$~~$  0.10 &$~~$  10.9 &$~~$ 5.79 &$~$  6.60 &$~$ 5.32 &$~~$ 12.0 &$~~$ 1.14 &$~$ 0.92 &$~$ 2.07 &$~$ \\
Run $B_3$ &$~~$  0.30 &$~~$  30.7 &$~~$ 9.72 &$~$  16.0 &$~$ 10.0 &$~~$ 12.9 &$~~$ 1.65 &$~$ 1.02 &$~$ 1.32 &$~$ \\
Run $B_4$ &$~~$  1.00 &$~~$  100. &$~~$ 17.5 &$~$  --- &$~$ --- &$~~$ --- &$~~$ --- &$~$ --- &$~$ --- &$~$ \\
Run $D_3$ &$~~$  0.30 &$~~$  30.7 &$~~$ 9.72 &$~$  9.94 &$~$ 8.54 &$~~$ 10.3 &$~~$ 1.02 &$~$ 0.88 &$~$ 1.06 &$~$ \\

\tableline
\end{tabular}
\tablenotetext{a}{The volume filling factor of the prominence}
\tablenotetext{b}{Volume averaged density of the prominence}
\tablenotetext{c}{Shock sweeping time scale}
\tablenotetext{d}{Damping time scale for $\bf{ \delta v_x}$}
\tablenotetext{e}{Damping time scale for $\bf{ \delta v_y}$}
\tablenotetext{f}{Damping time scale for $\bf{ \delta v_z}$}
\tablenotetext{g}{Damping time scale for $\bf{ \delta v_x}$ in unit of shock sweeping timescale}
\tablenotetext{h}{Damping time scale for $\bf{ \delta v_y}$ in unit of shock sweeping timescale}
\tablenotetext{i}{Damping time scale for $\bf{ \delta v_z}$ in unit of shock sweeping timescale}
\end{table}

\clearpage

\begin{table}
\caption{Estimated coronal shock parameters \label{tbl-1}}
\begin{tabular}{crrrrrrrrrrr}
\tableline
\tableline
\multicolumn{1}{c}{}& \multicolumn{1}{c}{$\beta$\tablenotemark{a}} &\multicolumn{1}{c}{$~~f_V$\tablenotemark{b}} &
 \multicolumn{1}{c}{$r\tablenotemark{c}$} & \multicolumn{1}{c}{$w/R_s$\tablenotemark{d}} &
\multicolumn{1}{c}{$~~M_f$\tablenotemark{e}} &
\multicolumn{1}{c}{$V_{sh,cor}$ \tablenotemark{f}}\\
\tableline
$\phi = 0^{\circ}$ &$~~$ 0.33 &$~~~$ 0.01 &$~~~$  1.17 &$~~~$  0.16 &$~~~$ 1.12 &$~~$ 62 km s$^{-1}$&$~~~~$ \\
$\phi = 45^{\circ}$ &$~~$ 0.15 &$~~~$ 0.01 &$~~~$  1.17 &$~~~$  0.22 &$~~~$ 1.13 &$~~$ 88 km s$^{-1}$&$~~~~$ \\

\tableline
\end{tabular}
\tablenotetext{a}{Plasma beta in the shock upstream corona}
\tablenotetext{b}{Volume filling factor of the prominence}
\tablenotetext{c}{Compression ratio of the coronal shock}
\tablenotetext{d}{Wave packet width in the corona in unit of a solar radius}
\tablenotetext{e}{Fast mode mach number of the coronal shock wave}
\tablenotetext{f}{Plasma velocity amplitude of the coronal shock wave}
\end{table}

\clearpage

\begin{figure}
\epsscale{.90}
\includegraphics[width=1.0\textwidth]{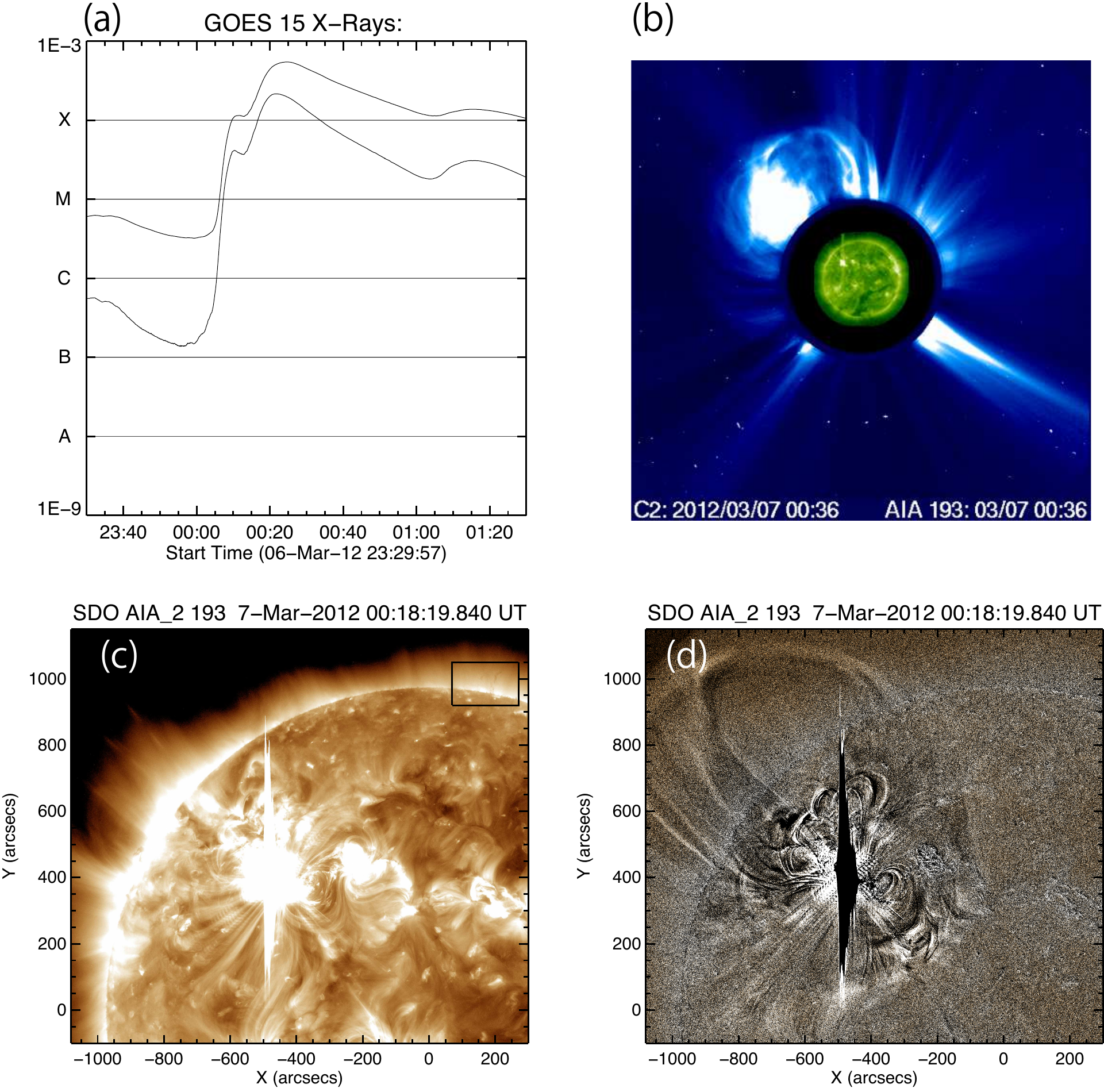}
\caption{(a) GOES Soft X-ray light curve. X5.4 class flare occurred at 00:04UT and peaked at 00:24UT on 7 March, 2012. (b) Composite of coronagraph image obtained with {\it SOHO}/LASCOC2 and 193 ${\AA}$ passband image by {\it SDO}/AIA. We can see the shock front surrounding the CME ejecta.
(c) {\it SDO}/AIA 193 ${\AA}$ band image showing the emission from 1MK coronal plasma at 00:18:19UT. The pixels around flaring active region AR11429 are saturated. The black rectangle 'R1' shows the FOV of Figure 3 (a) - (h).
(d) Difference image made from two successive snapshots of AIA 193 ${\AA}$ passband taken at 00:18:19UT and 00:18:07UT. A dome-like disturbance expanding above AR11429 is clearly seen.
\label{flare}}
\end{figure}
\begin{figure}
\includegraphics[width=1.0\textwidth]{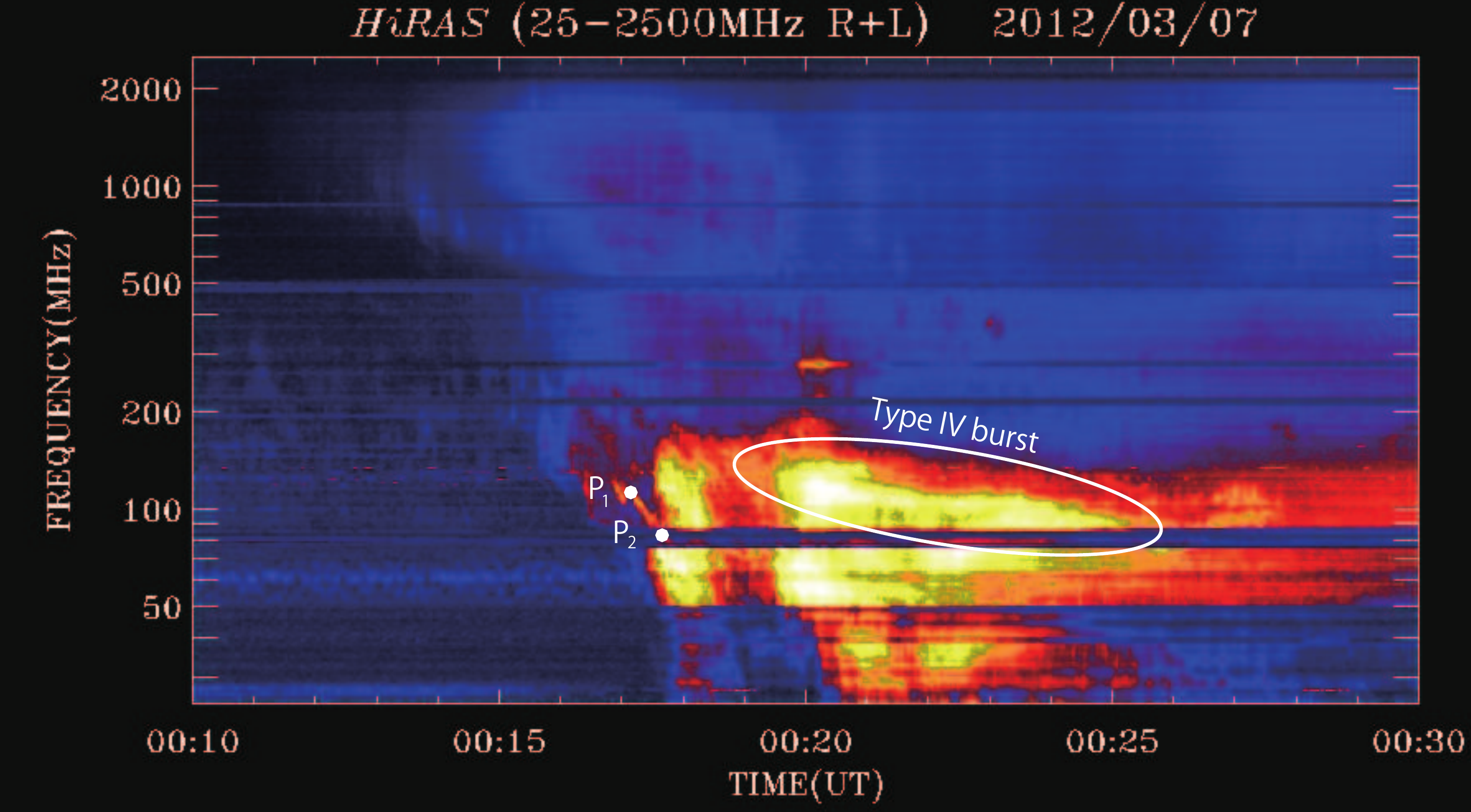}
\caption{Radio burst associated with the X5.4 flare in 7 March, 2012. We can see a clear linear structure showing the characteristic signature of Type II burst which drifted from $f=112$ MHz at 00:17:10UT (indicated as $P_1$) to $f=88$ MHz at 00:17:38UT (indicated as $P_2$). Type IV burst is also seen as indicated in the figure.
\label{flare}}
\end{figure}

\begin{figure}
\centering
\includegraphics[width=0.6\textwidth]{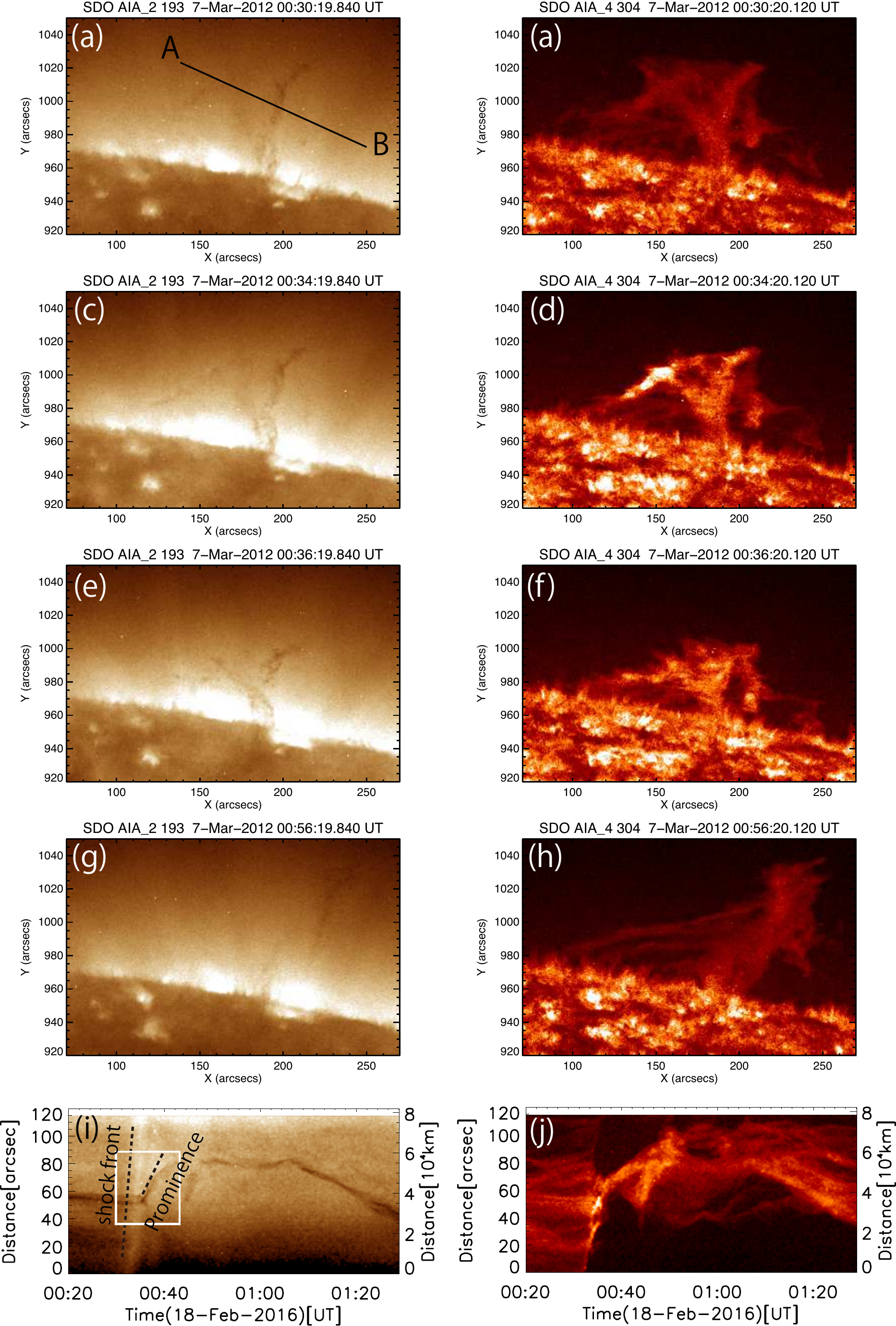}
\caption{Time evolution of prominence activation by coronal shock wave seen in AIA193 and 304 {\AA} bands. Panels (a) and (b) show the prominence just before coronal shock front reached the prominence. After the coronal shock wave arrive at the prominence (panels (c) and (d)), they propagate further accelerating the prominence in the direction of propagation (panels (e) and (f)). During acceleration, the prominence strongly brighten in AIA 304 {\AA} images. Panels (g) and (h) show the prominence with its maximum displacement from the original position seen in (a) and (b). (i) and (j) show time-distance plot along cut AB shown in (a) from 00:20UT to 01:30UT. The propagating shock front is seen as a bright linear feature. When the shock front hit the prominence (which is seen as dark trajectory in panel (i)), it is suddenly accelerated. After the shock have passed, the prominence threads moves in somewhat disordered manner as seen in panel (j). 
\label{flare}}
\end{figure}

\begin{figure}
\includegraphics[width=1.0\textwidth]{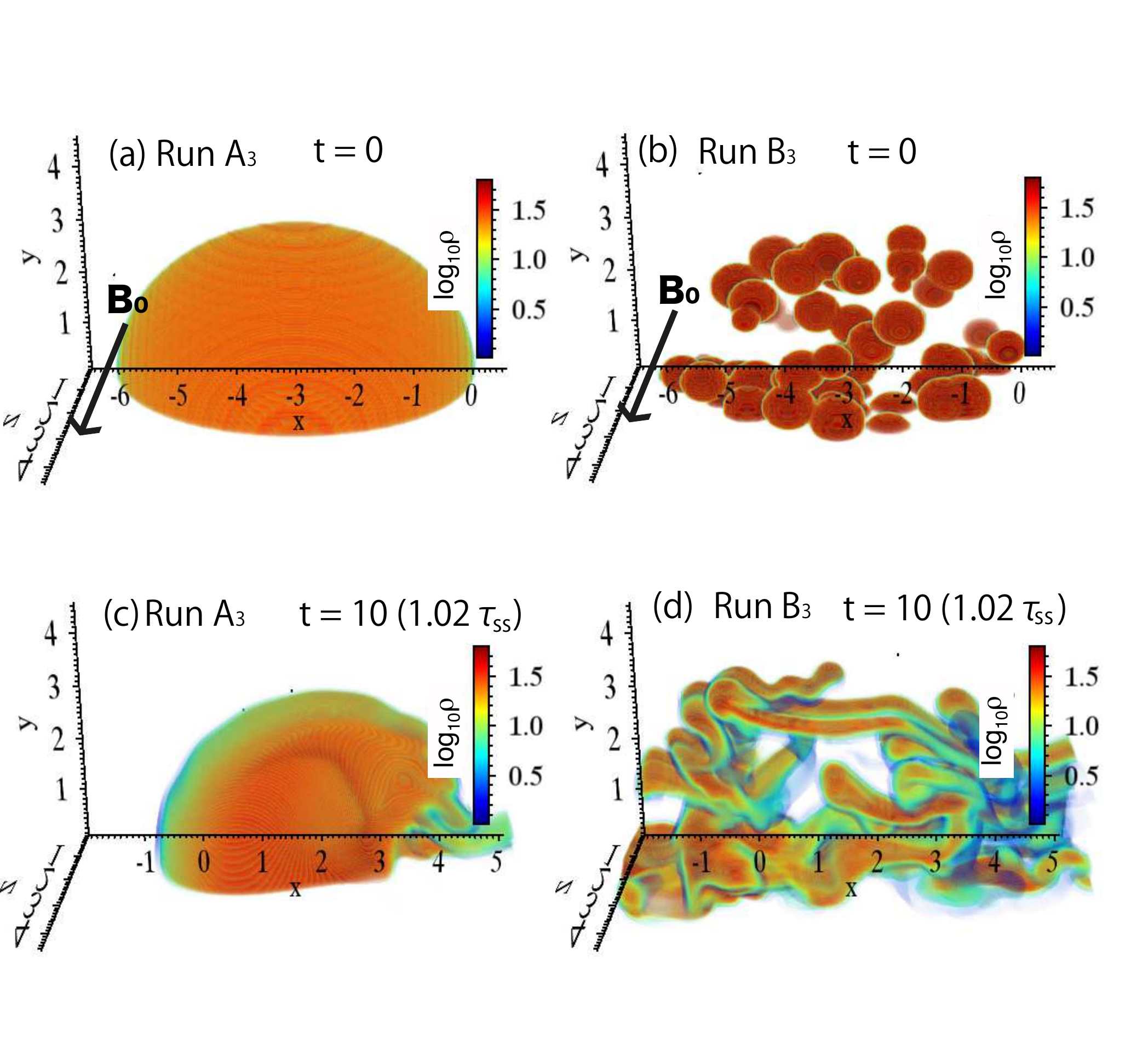}
\caption{Volume rendering of density in logarithmic scale in simulation Runs $A_3$ (panels (a) and (c)) and $B_3$ (panels (b) and (d)). Panels (a) and (b) show initial ($t = 0$) density distribution, while panels (c) and (d) shows the one at $t = 10$($=1.02\tau_{ss}$). We note that the range of the x-axis of the plots in panels (a) and (b) is between -6.5 and 0.5, while that in panels (c) and (d) is between -2 and 5. The values in the color bars corresponds to the logarithm of prominence density. The direction of initial magnetic field $\bf{B_0}$ is also shown in panels (a) and (b).
\label{flare}}
\end{figure}

\begin{figure}
\includegraphics[width=1.0\textwidth]{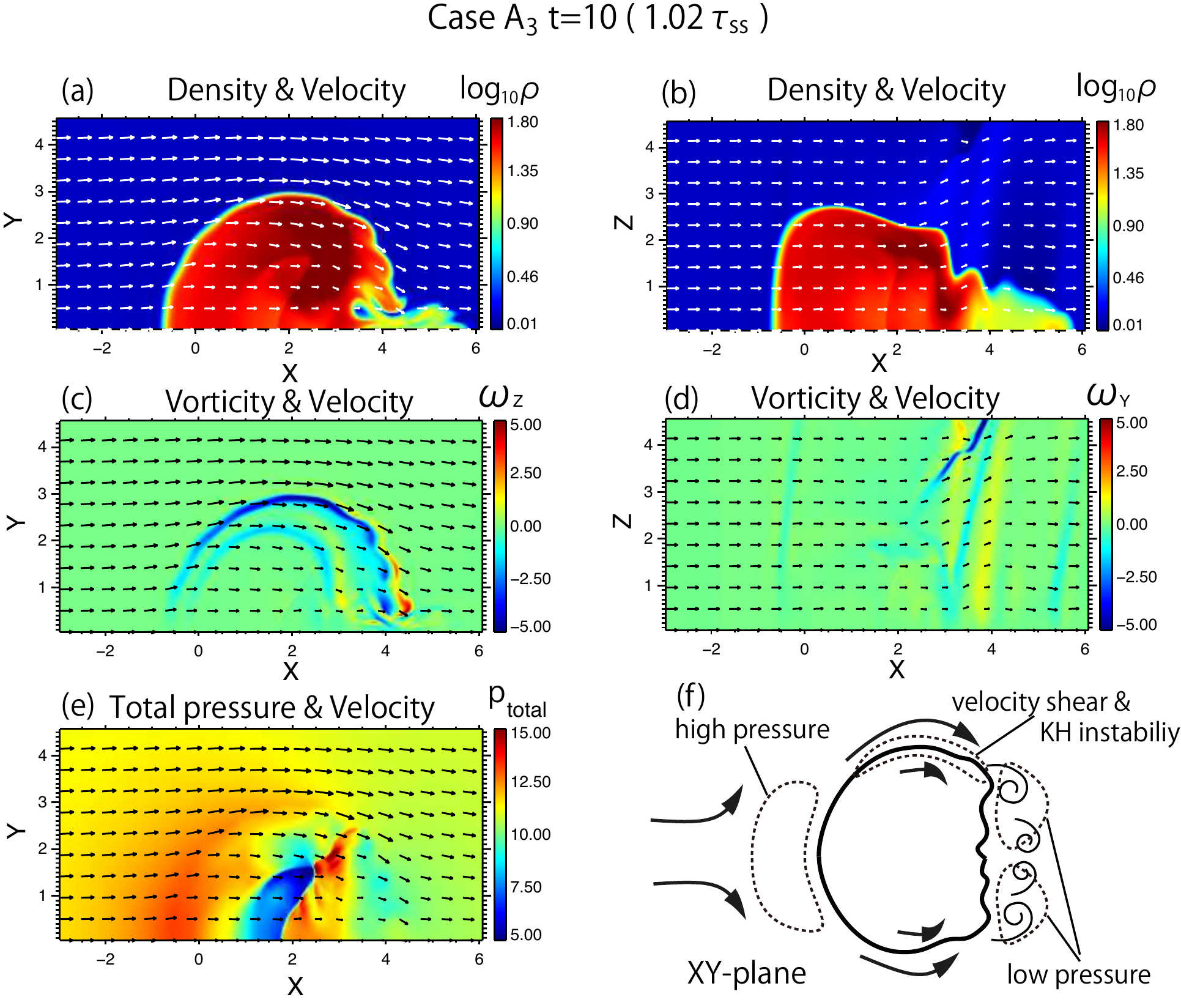}
\caption{Density, vorticity and total pressures in runs $A_3$ at $t=10$. Arrows represent velocity fields in the plane of the plot. Logarithmic density distributions in XY- or XZ- planes are shown with color contours in panels (a) and (b), respectively. In panel (c) and (d), vorticity components perpendicular to XY- or XZ- planes are shown with color contours, respectively. In panel (e), total plasma pressure (magnetic pressure$+$gas pressure) in XY plane is shown. Panel (f) is the schematic figure shows how Kelvin-Helmholtz instability evolves along the velocity shear layer in XY-plane. The resultant formation of low pressure region behind the cloud is also indicated.
\label{flare}}
\end{figure}

\begin{figure}
\includegraphics[width=1.0\textwidth]{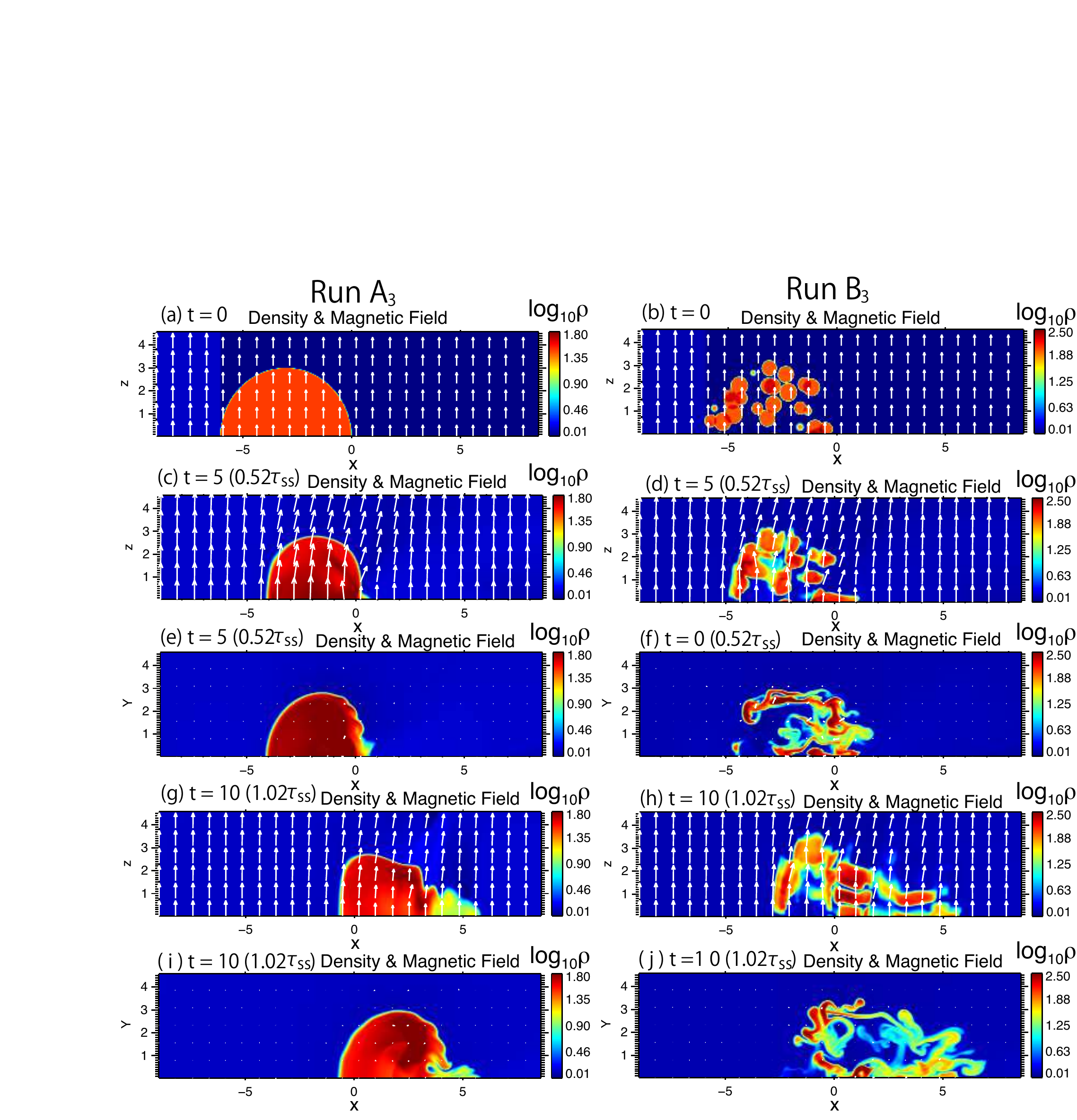}
\caption{Two dimensional cut for the density distribution in Runs $A_3$(left column) and  $B_3$ (right column). The density distributions are shown in logarithmic scale with color contours. White arrows in each panels show the magnetic field vector in the plane of the plots. The panels in third and fifth rows are plots in $z=0$ plane, while other panels shows plots in $y=0$ plane. Simulation time for each snapshot in unit of shock sweeping time scale $\tau_{ss}$ is shown in brackets.
\label{flare}}
\end{figure}

\begin{figure}
\includegraphics[width=1.0\textwidth]{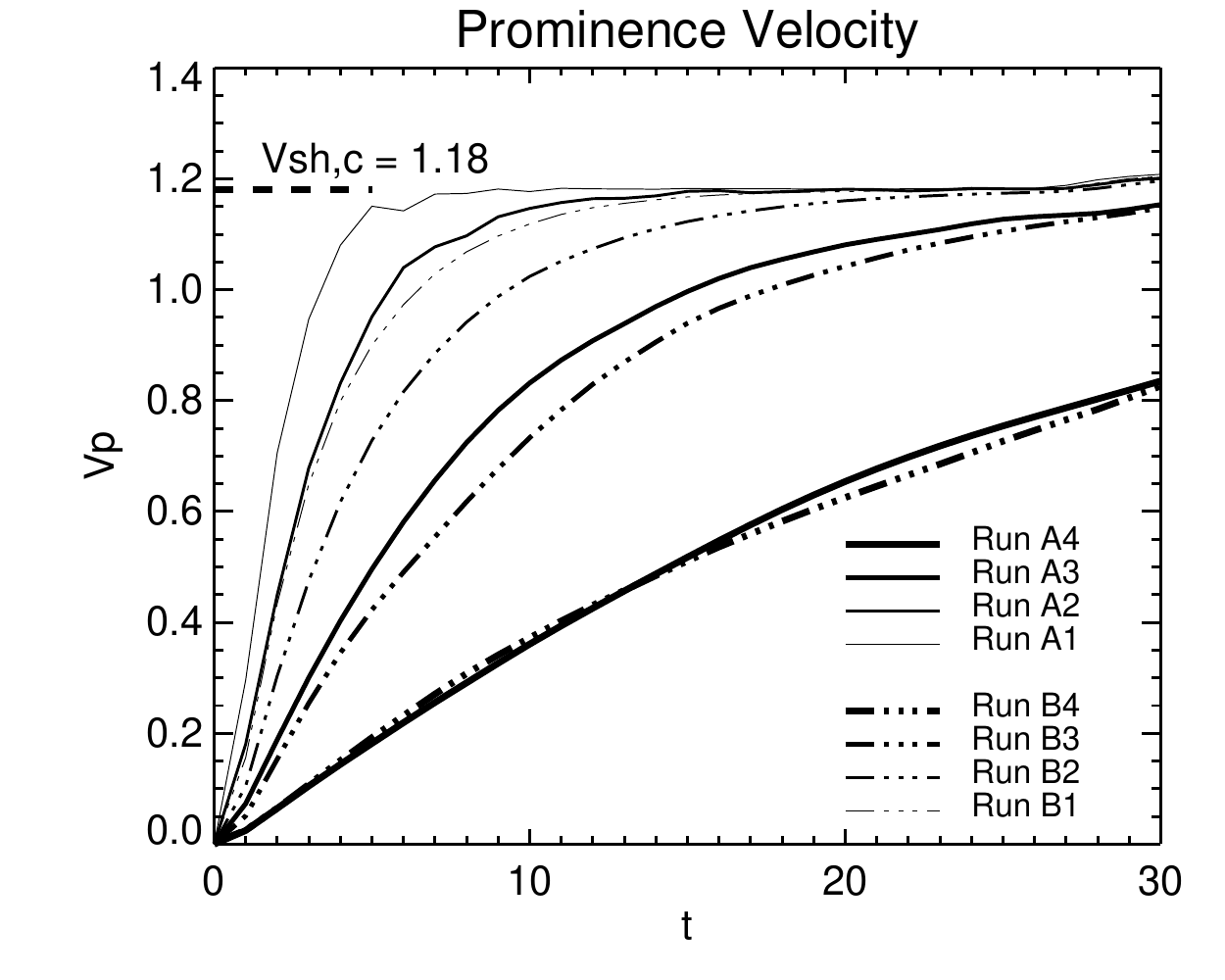}
\caption{Time evolution of prominence velocity $V_p$ in Runs $A_i$(solid lines) and $B_i$ (dash dotted lines). Thicker lines represent heavier prominence (i.e. larger $f_V$).
\label{flare}}
\end{figure}

\begin{figure}
\includegraphics[width=1.0\textwidth]{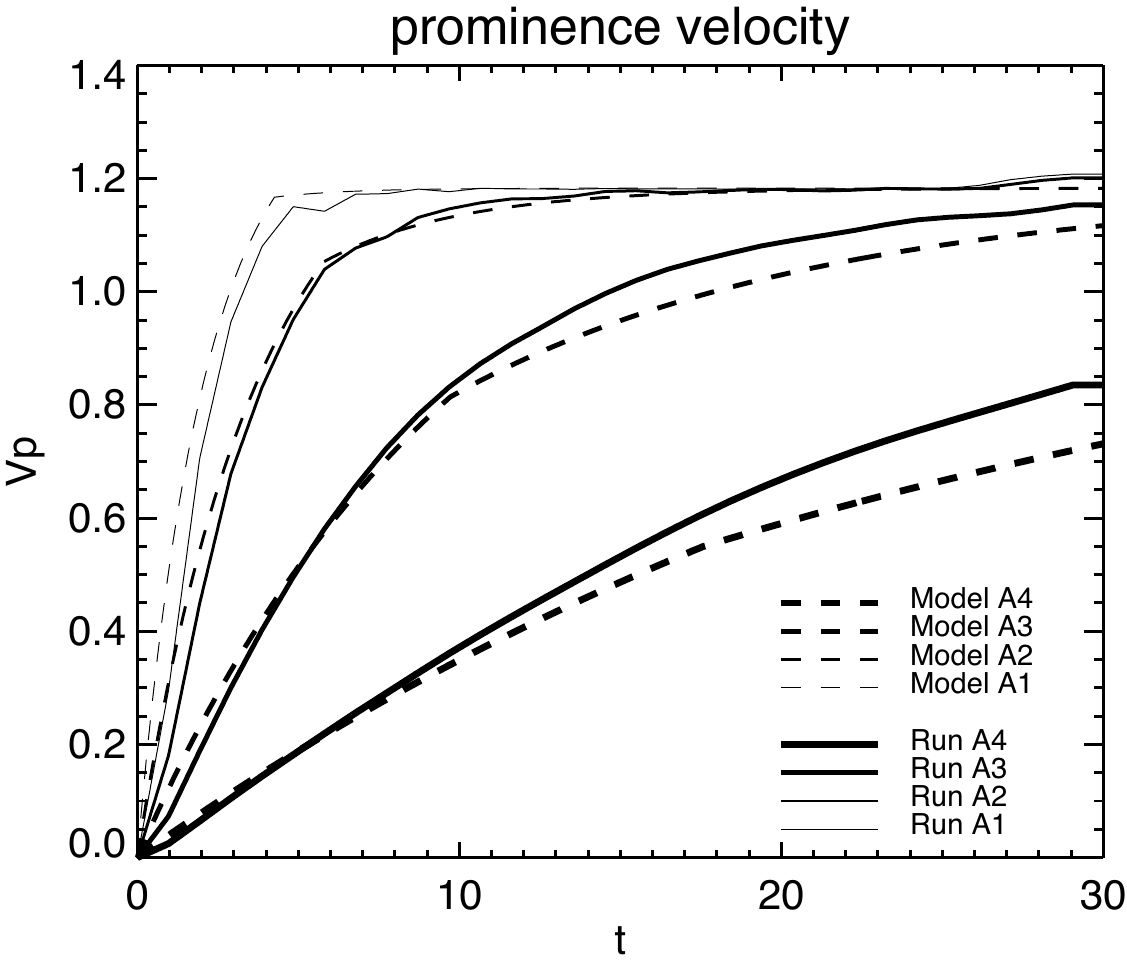}
\caption{Comparizon between the time evolution of $V_p$ of simulation Runs $A_i$ (solid lines) and that of phenomenological Models $A_i$ (dashed lines). Thicker lines represent heavier prominence (i.e. larger $f_V$).
\label{flare}}
\end{figure}

\begin{figure}
\includegraphics[width=1.0\textwidth]{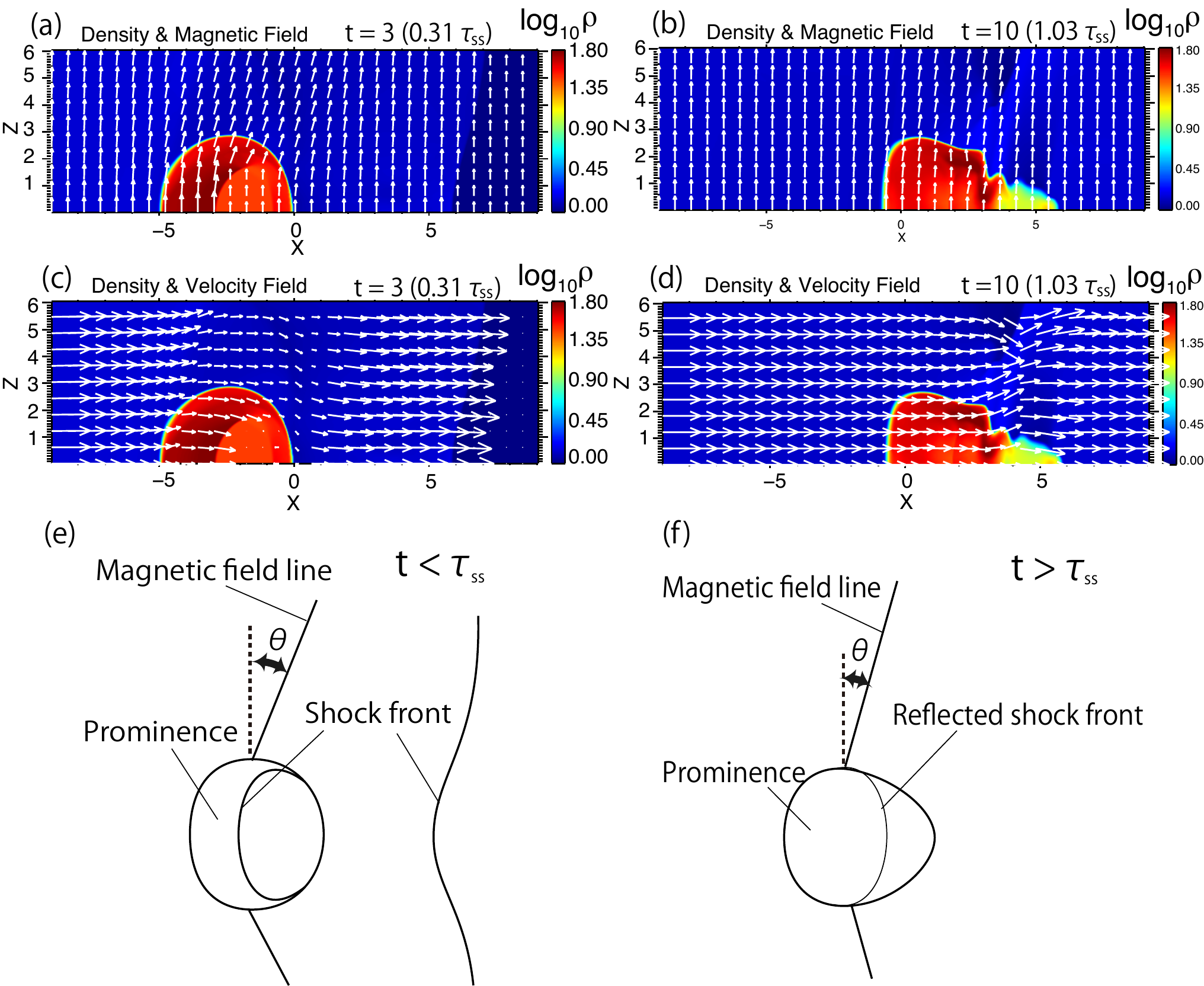}
\caption{Comparison between simulation Runs $A_i$ and phenomenological Models $A_i$. Top two rows shows density distribution of Run $A_3$ in logarithmic scale with color contours. Arrows in panels (a) and (b) show magnetic fields and those in panels (c) and (d) show velocity fields, both in XZ plane. Panels (e) and (f) are schematic figures of phenomenological prominence activation model during ($t < \tau_{ss}$) and after ($t > \tau_{ss}$) the shock sweeping acceleration works. The coronal shock front that passed the cloud is distorted as shown in (e), while the shock front transmitted into the prominence material is reflected back and forth at prominence-corona boundary as shown in (e) and (f).
\label{flare}}
\end{figure}

\begin{figure}
\includegraphics[width=1.0\textwidth]{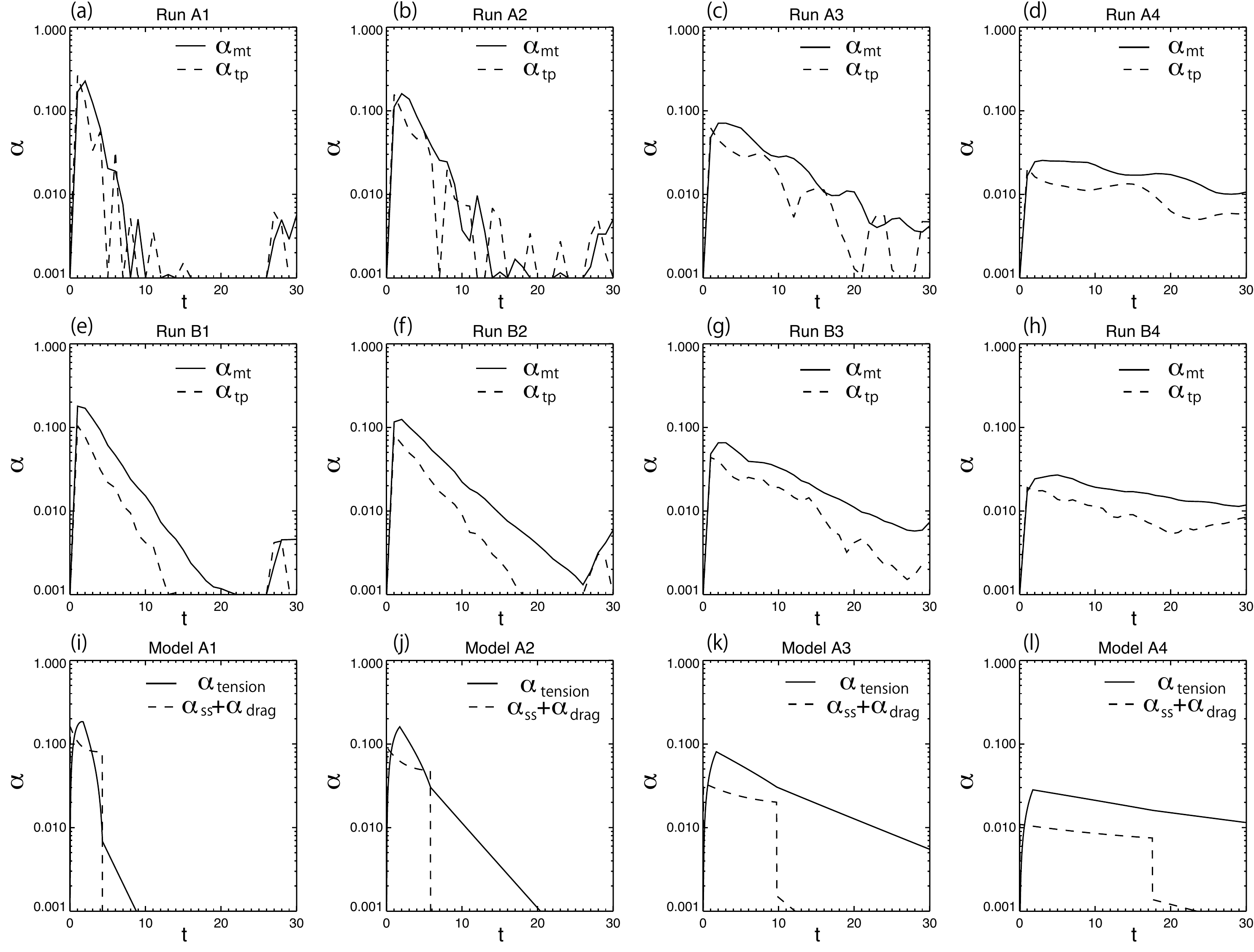}
\caption{Time evolution of prominence accelerations. In the top two rows, total pressure gradient force acceleration $\alpha_{tp}$ and magnetic tension force acceleration $\alpha_{mt}$ in simulation Runs $A_i$ and $B_i$ are shown as dashed and solid lines, respectively. In the bottom row, summation of shock sweeping acceleration and fluid drag acceleration $\alpha_{ss}+\alpha_{drag}$ and magnetic tension force acceleration $\alpha_{tension}$ in Models $A_i$ are shown as dashed and solid lines, respectively.
\label{flare}}
\end{figure}

\begin{figure}
\includegraphics[width=1.0\textwidth]{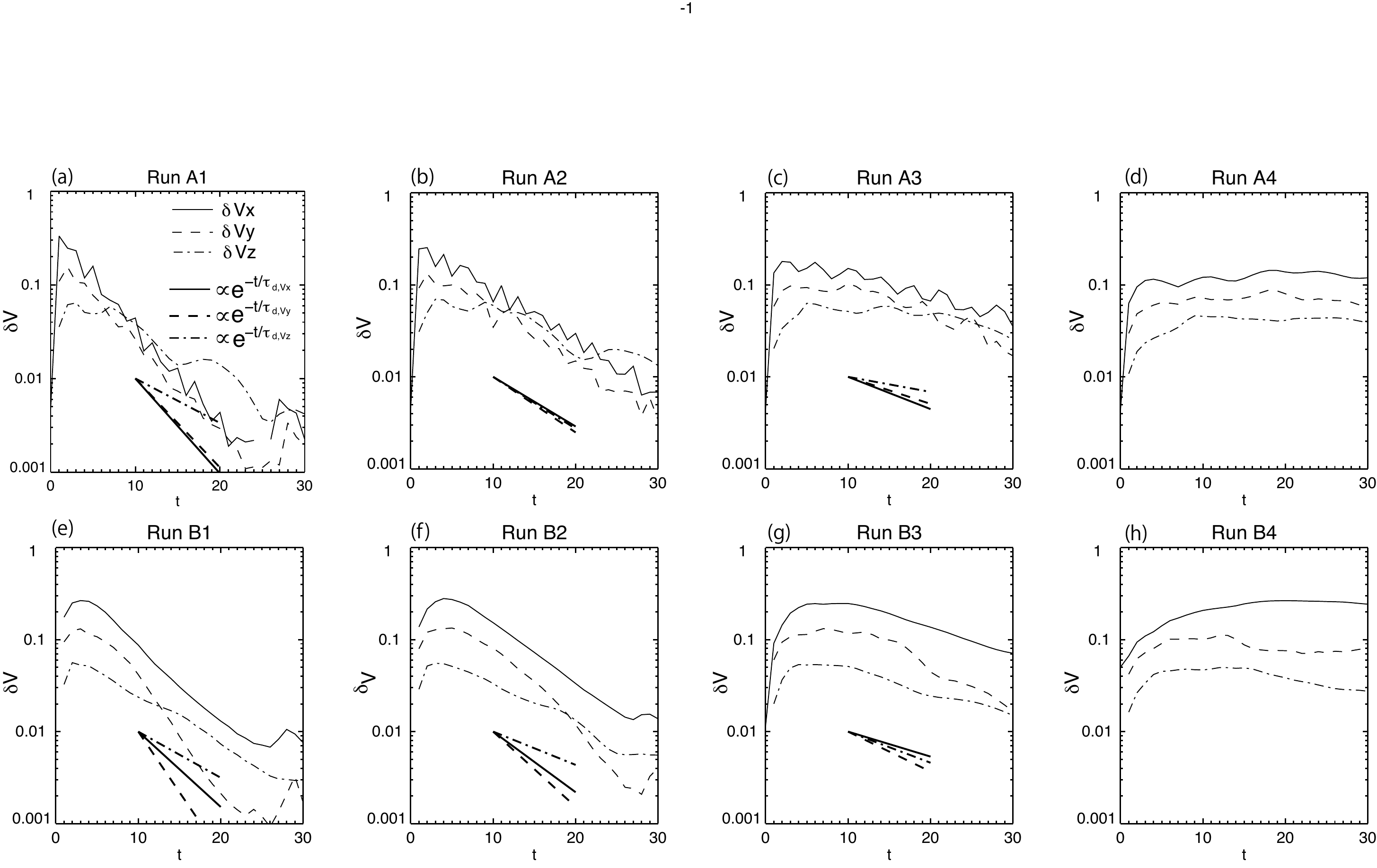}
\caption{Time evolution of velocity dispersion $\delta V_x$ (solid lines), $\delta V_y$ (dashed lines) and $\delta V_z$ (dash-dotted lines) in Runs $A_i$ and $B_i$. Exponential damping timescales are estimated for the three variables in each runs by least square fit during the time between $t=10$ and $t=20$ in Runs $A_i$ and $B_i$ ($i$ is from 1 to 3). The resultant slopes for exponential damping are shown as thick lines for each variables. Estimated exponential damping time scales $\tau_{d,vx}$, $\tau_{d,vy}$ and $\tau_{d,vz}$ are listed in table 1.
\label{flare}}
\end{figure}

\begin{figure}
\includegraphics[width=1.0\textwidth]{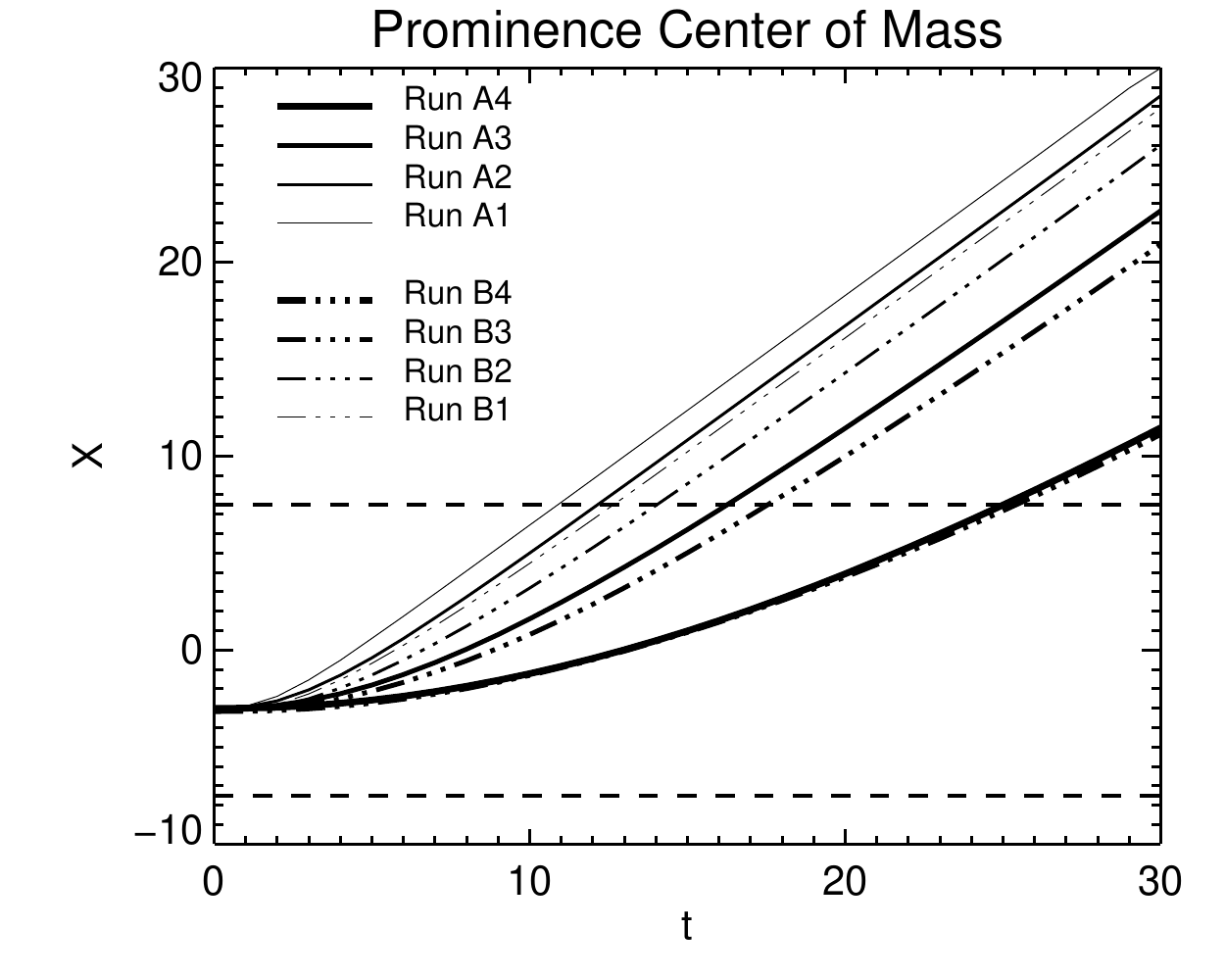}
\caption{Time evolution of x-coordinate of prominence center of mass in Runs $A_i$ (solid lines) and Runs $B_i$ (dash dotted lines). The boundary between uniform and non-uniform numerical grid is located at $X = \pm 7.5$ and expressed as dashed lines.
\label{flare}}
\end{figure}

\begin{figure}
\centering
\includegraphics[width=0.8\textwidth]{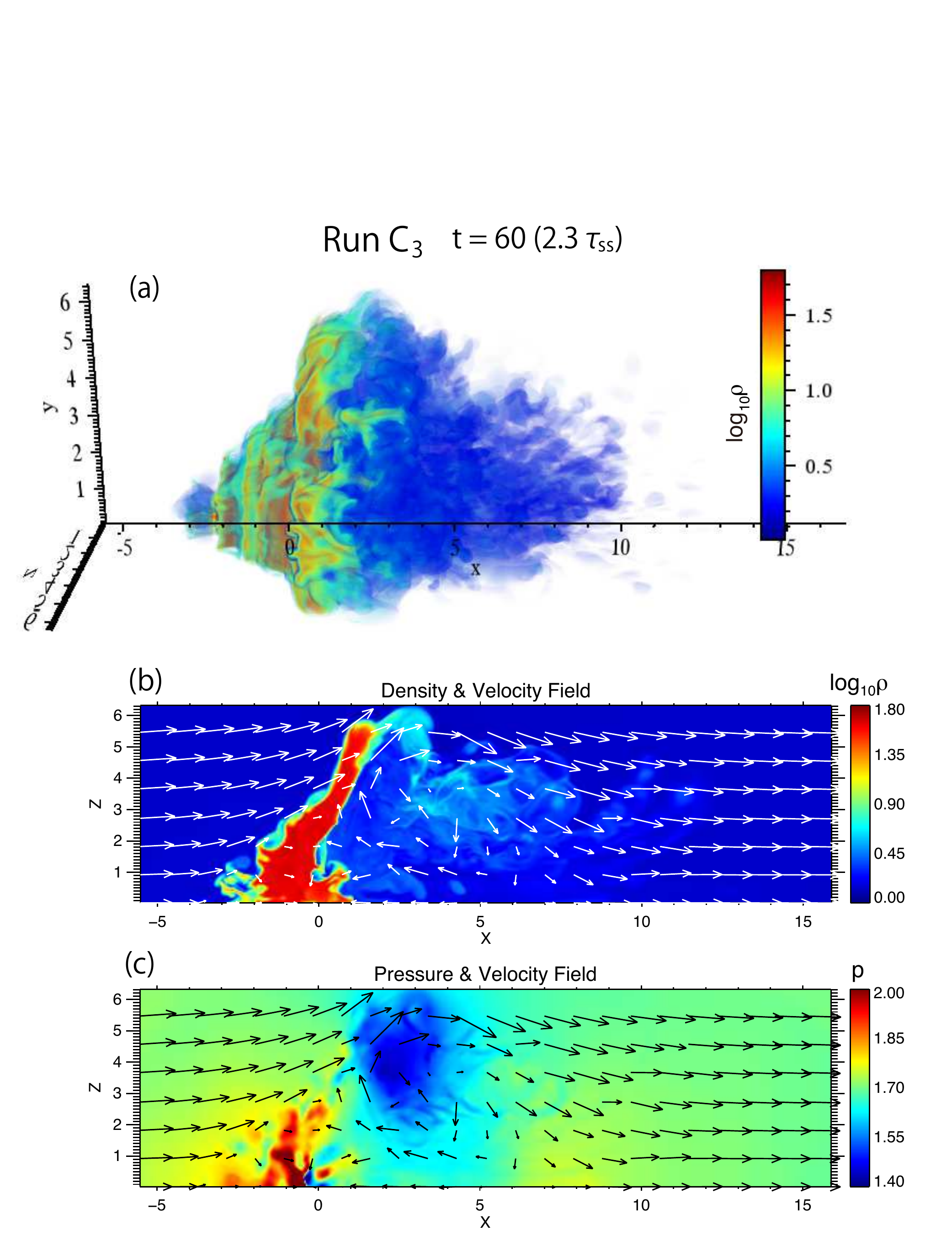}
\caption{Panel (a) shows the volume rendering of density distribution in simulation Run $C_3$ at $t = 60$. The values in the color bars corresponds to logarithm of density. Panel (b) and (c) show the density (in logarithmic scale) and pressure distributions in $y=0$ plane in simulation Run $C_3$ at $t = 60$, respectively. The arrows in panels (b) and (c) shows velocity field in $y=0$ plane.
\label{flare}}
\end{figure}

\begin{figure}
\includegraphics[width=1.0\textwidth]{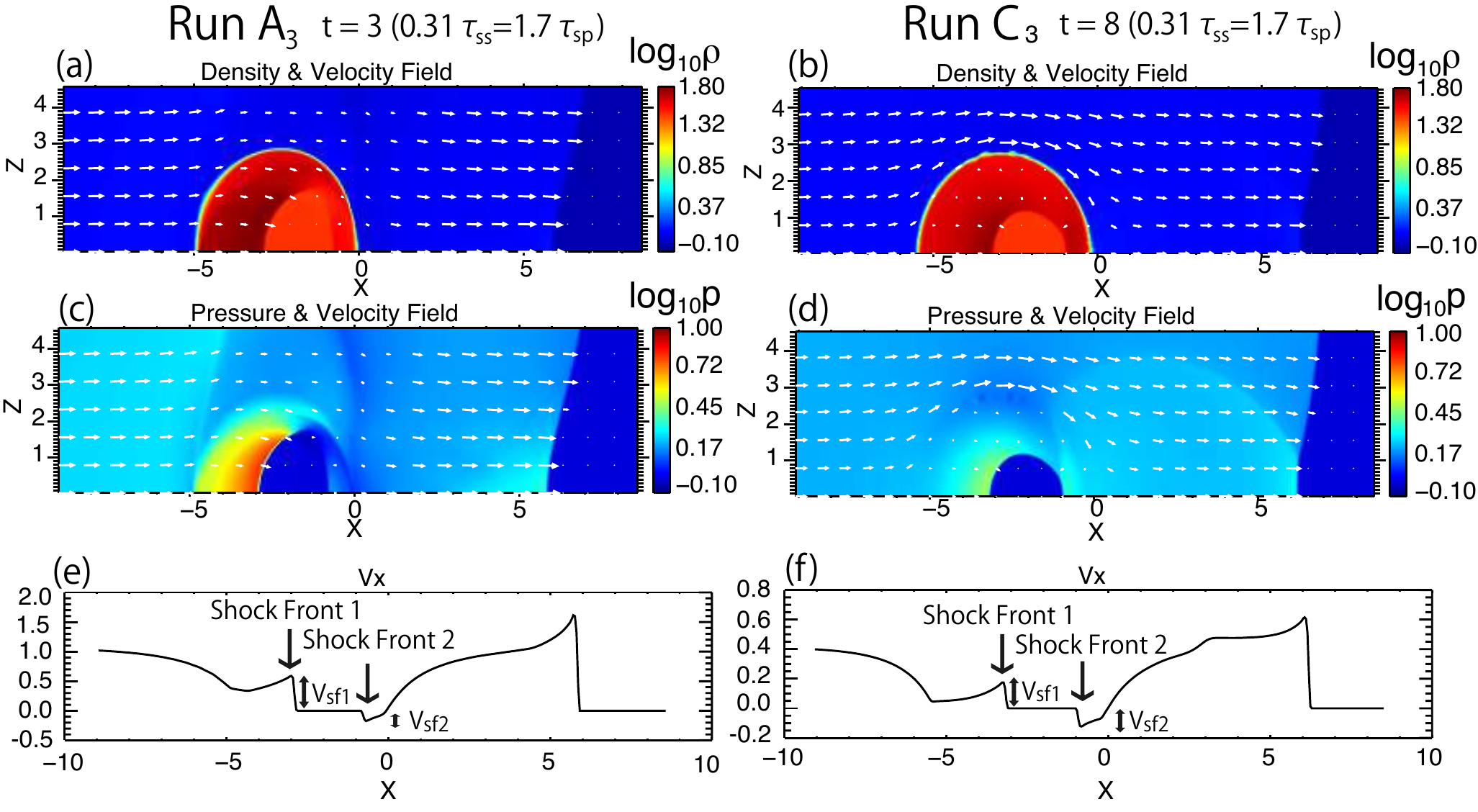}
\caption{Panels (a) and (c) show density and pressure distribution in simulation Run $A_3$ at $t = 3$ ($= 0.31\tau_{ss}=1.7\tau_{sp}$). Both density and pressure distribution is shown with color contour in logarithmic scale. In-plane velocity field is shown with white arrows in each panels. Panels (b) and (d) are those of Run $C_3$ at $t = 8$ ($= 0.31\tau_{ss} = 1.7\tau_{sp}$). Panel (e) and (f) show the distribution of $V_x$ along x-axis in Runs $A_3$ and $C_3$, respectively. The velocity amplitudes within prominence material for the shock fronts transmitted from ahead and behind of the prominence is indicated as $V_{sf1}$ and $V_{sf2}$, respectively.
\label{flare}}
\end{figure}
\clearpage

\begin{figure}
\epsscale{.90}
\includegraphics[width=1.0\textwidth]{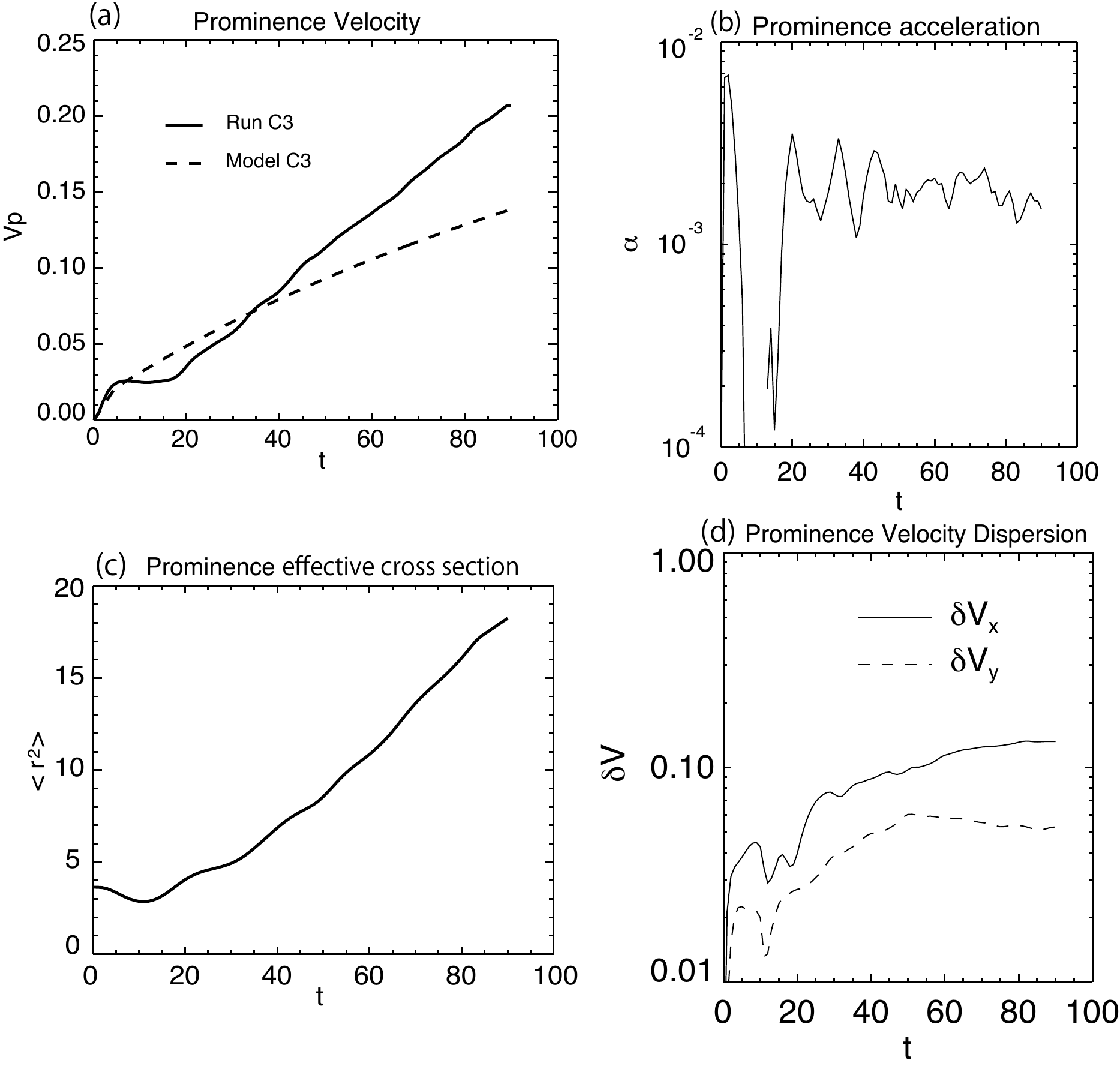}
\caption{Time evolution of (a): center of mass velocity $V_x$, (b): center of mass acceleration $\alpha$, (c): index of cross sectional area $<r^2>$ and (d): velocity dispersions $\delta V_x$ (solid line) and $\delta V_y$ (dashed line) of the cloud in Hydrodynamic Run $C_3$. In panel (a), prominence velocity expected from phenomenological model (dashed line) and the simulation result (solid line) are compared.
\label{flare}}
\end{figure}

\begin{figure}
\includegraphics[width=1.0\textwidth]{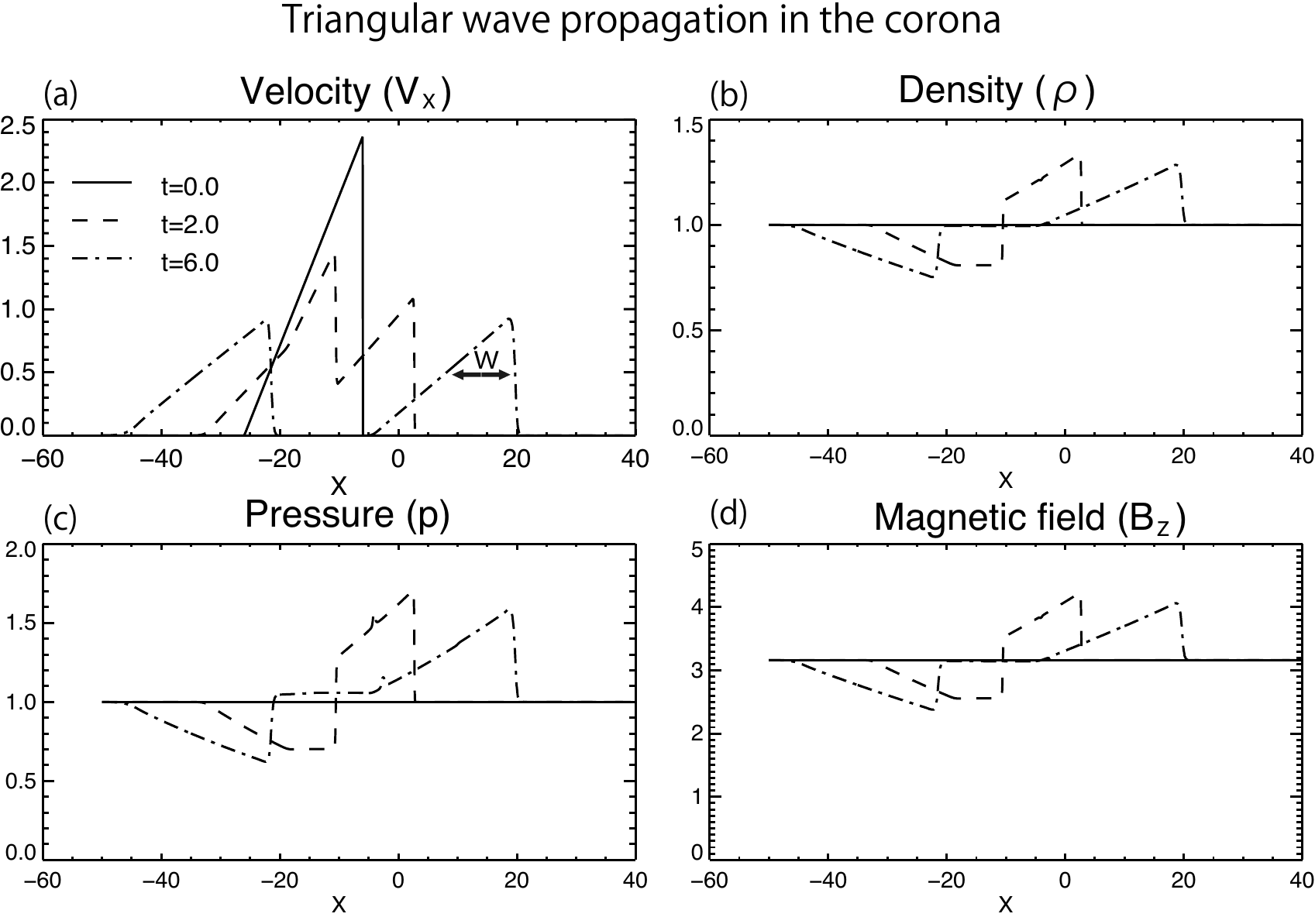}
\caption{1D MHD simulation results with the same initial condition with simulation Run $D_3$ without the prominence. We see the time evolution of wave packets that propagate to opposite directions with each other. The spacial distribution of $V_x$, $\rho$, $p$ and $B_z$ are shown in panels (a), (b), (c) and (d), respectively. The solid, dashed and dash-dotted lines in each panels represents physical values at $t=$0, 2 and 6, respectively.
\label{flare}}
\end{figure}

\begin{figure}
\includegraphics[width=1.0\textwidth]{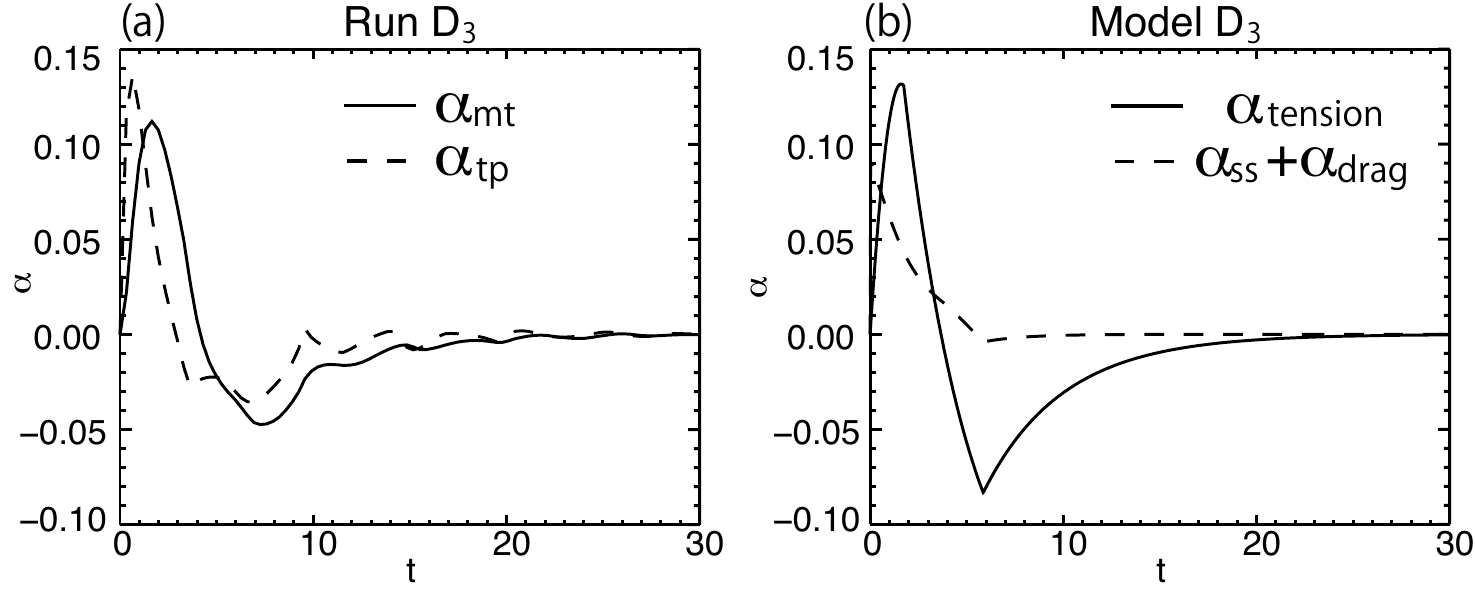}
\caption{The solid and dashed lines in panel (a) show prominence acceleration by magnetic tension force $\alpha_{mt}$ and that by total pressure gradient force $\alpha_{tp}$ in simulation Run $D_3$, respectively. The solid and dashed lines in panel (b) show prominence acceleration by magnetic tension mechanism $\alpha_{tension}$ and that both by shock sweeping and fluid drag mechanisms $\alpha_{ss}+\alpha_{drag}$ expected from phenomenological Model $D_3$, respectively.
\label{flare}}
\end{figure}

\begin{figure}
\includegraphics[width=1.0\textwidth]{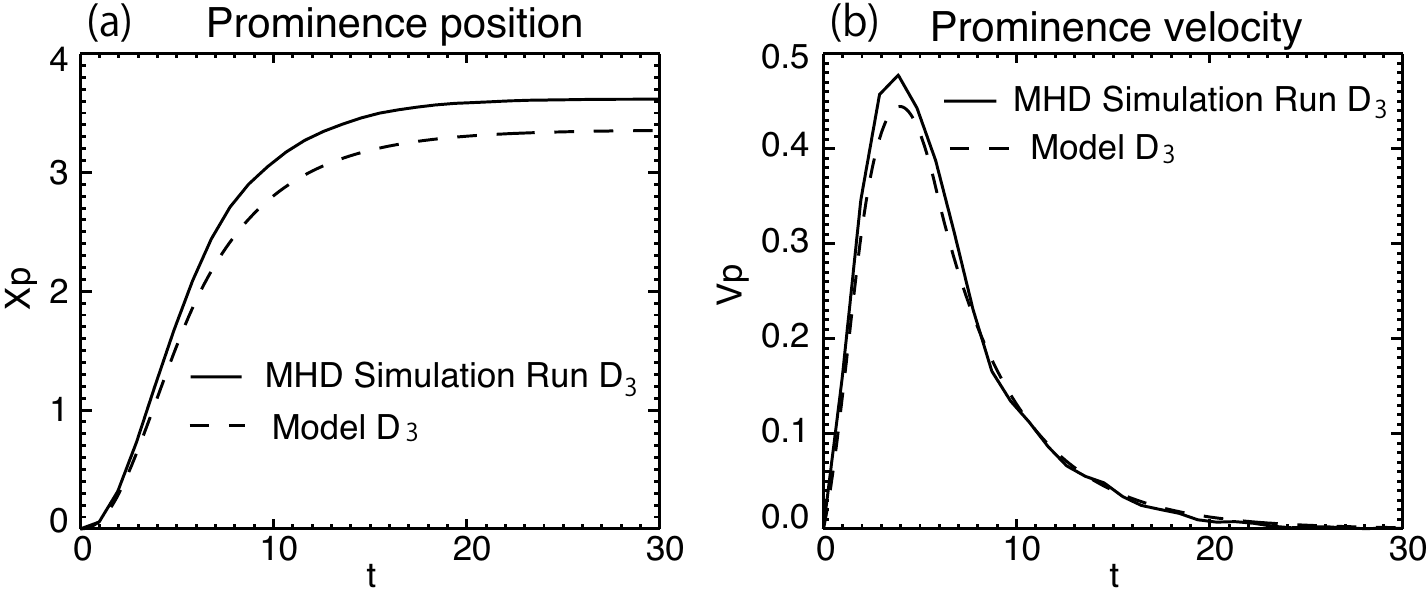}
\caption{Prominence center of mass position $X_p$ (panel (a)) and velocity $V_p$ (panel (b)) in simulation Run $D_3$ (solid lines) and phenomenological Model $D_3$ (dashed lines).
\label{flare}}
\end{figure}

\begin{figure}
\includegraphics[width=1.0\textwidth]{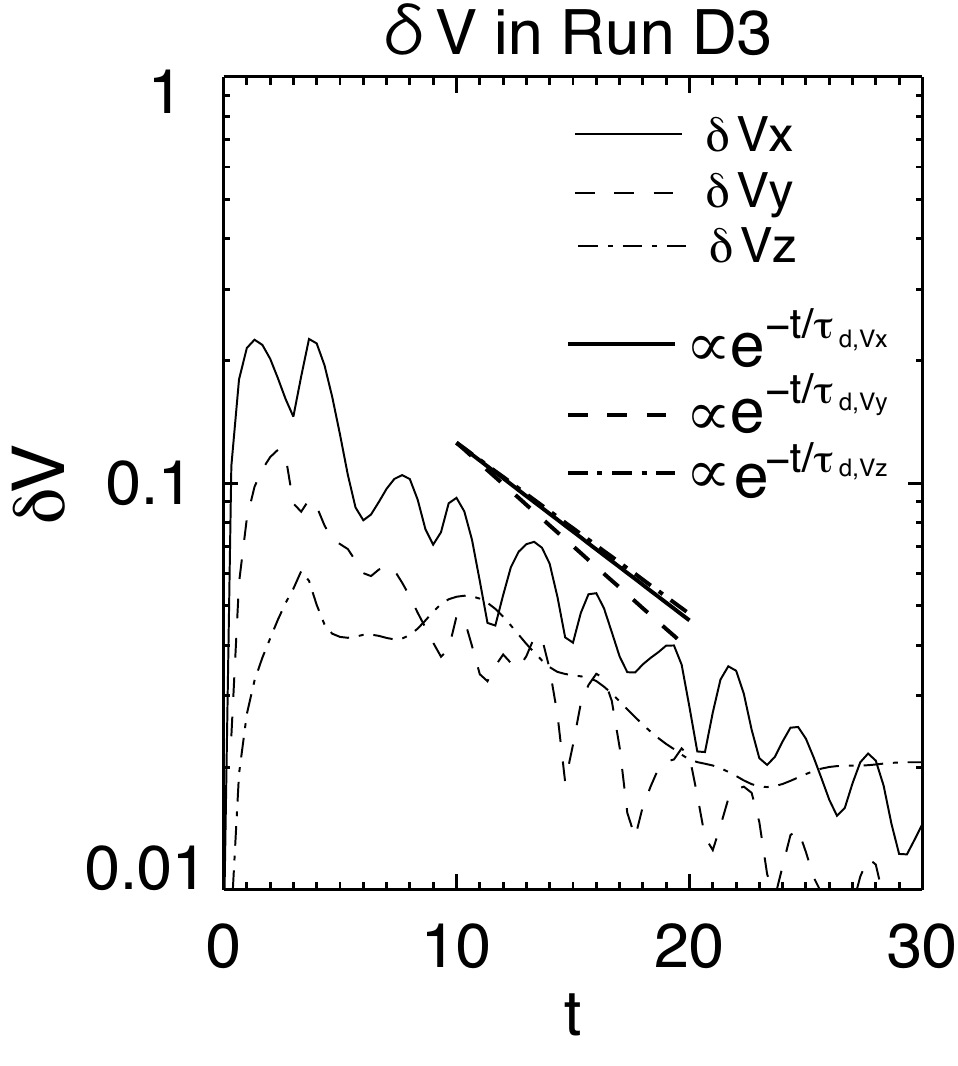}
\caption{The time evolution of velocity dispersion $\delta V_x$ (solid lines), $\delta V_y$ (dashed lines) and $\delta V_z$ in simulation Run $D_3$. Exponential damping timescales are estimated for the three variables in each runs by least square fit during the time between $t=10$ and $t=20$. The resultant slopes for exponential damping are shown as thick lines for each variables.
\label{flare}}
\end{figure}

\begin{figure}
\includegraphics[width=1.0\textwidth]{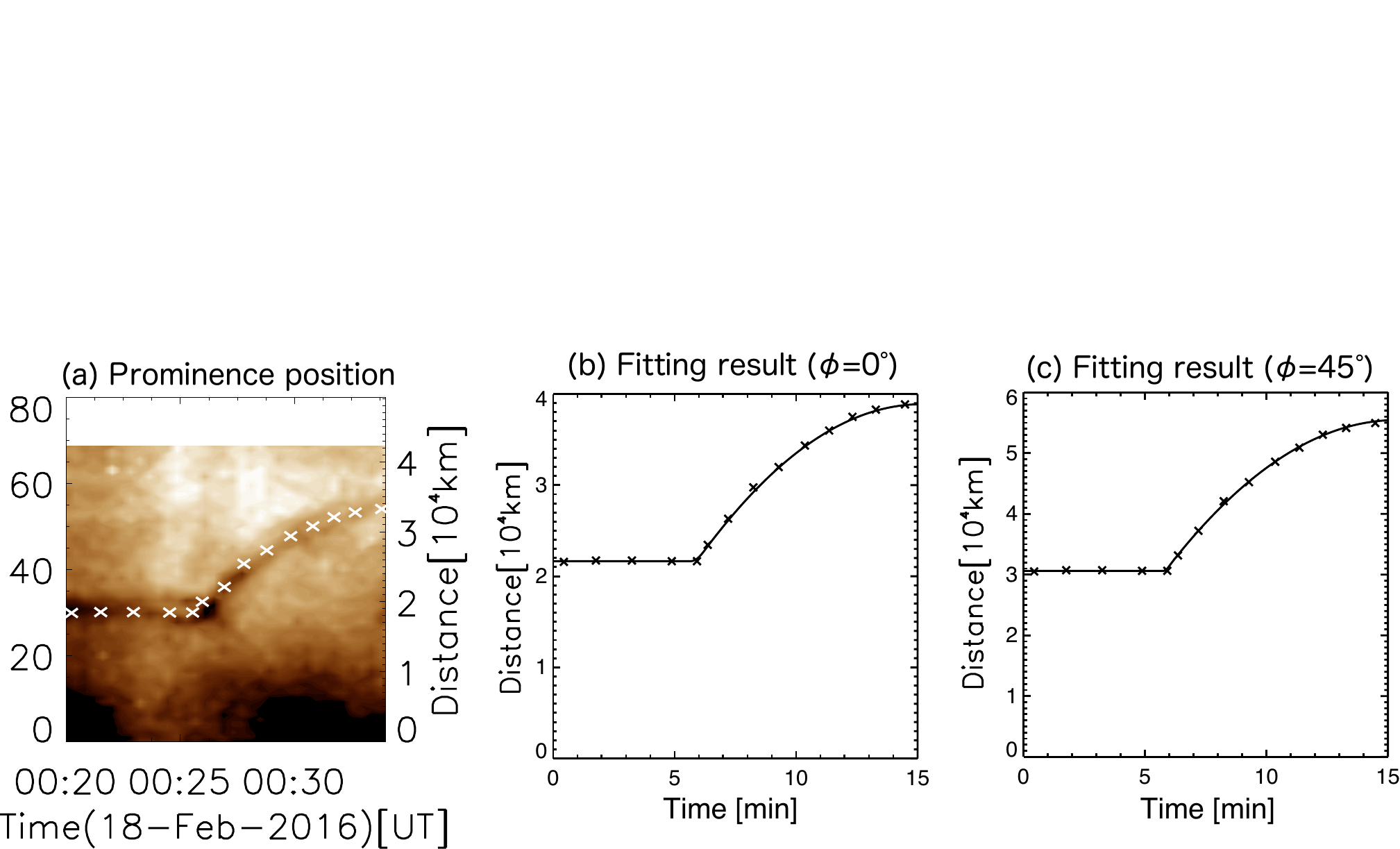}
\caption{Application of the phenomenological model to observation. Extracted prominence positions from time-distance diagram of prominence activation are shown as white crosses in panel (a). Panel (b) and (c) shows the fitting result by phenomenological model of triangular wave-prominence interaction in the cases of $\phi=0^{\circ}$ and $\phi=45^{\circ}$, respectively. Black crosses in panel (b) and (c) are the extracted position from observation, and the solid lines are the best-fit curves from the phenomenological model. The fitting result is listed in table 2.
\label{flare}}
\end{figure}

\end{document}